\documentclass[10pt,twoside]{article}

\usepackage{amssymb}
\usepackage{amsthm}
\usepackage{amsmath}

\usepackage{subfigure}
\usepackage{graphicx}

\renewcommand{\sectionmark}[1]%
        {\markboth%
            {{\sc Spherically symmetric relativistic stellar structures}}%
                {{\rm Section \thesection}\quad{\rm #1}}}

\newcommand{\proofend}{\hfill\rule{0.2cm}{0.2cm}}
\newcommand{\sfrac}[2]{{\textstyle{\frac{#1}{#2}}}}

\theoremstyle{plain}
\newtheorem{theorem}{Theorem}[section]

\theoremstyle{remark}

\newtheorem*{example}{Example}
\newtheorem*{examples}{Examples}
\newtheorem*{remark}{Remark}

\newtheorem*{assumptions}{Assumptions}

\begin{document}

\title{\sc Spherically symmetric relativistic stellar structures}
\author{\sc
J.\ Mark Heinzle$^{1}$\thanks{Electronic address: {\tt
mheinzle@thp.univie.ac.at}}\ ,\
Niklas R\"ohr$^{2}$\thanks{Electronic address:
{\tt Niklas.Rohr@kau.se}}\ ,\
\ and
Claes Uggla$^{2}$\thanks{Electronic address:
{\tt Claes.Uggla@kau.se}}\\
$^{1}${\small\em Institute for Theoretical Physics, University of Vienna}\\
{\small\em A-1090 Vienna, Austria}\\
$^{2}${\small\em Department of Physics, University of Karlstad}\\
{\small\em S-651 88 Karlstad, Sweden}}

 \maketitle

\begin{abstract}

We investigate relativistic spherically symmetric static perfect
fluid models in the framework of the theory of dynamical systems.
The field equations are recast into a regular dynamical system on
a 3-dimensional compact state space, thereby avoiding the
non-regularity problems associated with the
Tolman-Oppenheimer-Volkoff equation. The global picture of the
solution space thus obtained is used to derive qualitative
features and to prove theorems about mass-radius properties. The
perfect fluids we discuss are described by barotropic equations of
state that are asymptotically polytropic at low pressures and, for
certain applications, asymptotically linear at high pressures. We
employ dimensionless variables that are asymptotically homology
invariant in the low pressure regime, and thus we generalize
standard work on Newtonian polytropes to a relativistic setting
and to a much larger class of equations of state.
Our dynamical systems framework is particularly suited for numerical
computations, as illustrated by several numerical examples,
e.g., the ideal neutron gas and examples that involve phase transitions.

\end{abstract}
\bigskip
\centerline{\noindent PACS numbers:
04.40.Dg,  
04.40.Nr,  
97.10.--q, 
02.90.+p.} 

\vspace{10pt}


\vfill
\newpage


\section{Introduction}
\label{introduction}

The line element of a static spherically symmetric spacetime can
be written as
\begin{equation}\label{ds2}
ds^2= -c^2 e^{2\Phi(r)} dt^2 +  e^{2\Lambda(r)}dr^2 +
r^2(d{\theta}^2 + {\sin}^2{\theta}\, d{\phi}^2)\, ,
\end{equation}
where $r$ is the radial area coordinate and $c$ the speed of
light, which we have chosen to introduce explicitly.
A perfect fluid is described by the energy-momentum tensor
$T_{\mu\nu} = \rho u_\mu u_\nu + p (g_{\mu\nu} + u_\mu u_\nu)$,
where $\rho$ is the mass-energy density and $p$ the pressure as
measured in a local rest frame; compatibility with static symmetry
requires the unit vector $u^\mu$ to be parallel to the static
Killing vector, i.e., $u^\mu = c^{-1} e^{-\Phi} \delta^\mu_0$.
Einstein's field equations can be written so that the
``relativistic potential'' $\Phi$ is given by
\begin{equation}
\label{potential}
\frac{d\Phi}{dr} = -\frac{1}{p+\rho c^2} \,\frac{dp}{dr}\: ,
\end{equation}
while the remaining equations take the form
\begin{subequations}\label{oppvol}
\begin{align}
\label{mrov}
  \frac{dm}{dr} &= 4\pi r^2 \rho\, , \\
\label{oppvoleq}
  \frac{dp}{dr} &= -\frac{Gm\rho}{r^2}\left(1 + \frac{p}{\rho c^2}\right)
  \left(1 + \frac{4\pi r^3 p}{m c^2}\right)
  \left(1 - \frac{2G m}{c^2 r}\right)^{-1}\, ,
\end{align}
\end{subequations}
where $G$ is the gravitational constant and where the mass function
$m(r)$ is defined by $\exp(2\Lambda) = (1 - \frac{2G m(r)}{c^2 r})^{-1}$.
To form a determined system, these equations must be supplemented by equations
characterizing the properties of the perfect fluid matter, e.g., by
a barotropic equation of state $\rho(p)$.

Eqs.~(\ref{oppvol}) are the equations that govern hydrostatic equilibrium of
a self-gravitating spherically symmetric perfect fluid body,
Eq.~(\ref{oppvoleq}) is the famous Tolman-Oppenheimer-Volkoff (TOV) equation
\cite{tol39, oppvol39}.
The TOV equation is fundamental to the study of relativistic stellar models,
nonetheless it is associated with certain problems.
One is that the equation is singular at $r=0$, which leads to
mathematical complications. It was only recently that
such central issues as existence and uniqueness of
regular static spherically symmetric perfect
fluid models could be established \cite{Rendall/Schmidt:1991}.
A less known but perhaps even more problematic feature is that
the right hand sides of Eqs.~(\ref{oppvol}) are not differentiable
in $p$ at $p=0$ for many barotropic equations of state, e.g.,
for asymptotically polytropic equations of state, discussed below.
A similar problem occurs when $p\rightarrow \infty$; even though this
limit is unphysical, this is still important.
The relevance of the limit is due to the fact
that there are several physical problems that are directly or indirectly
governed by the limit $p\rightarrow \infty$.
Consider, for example, the Buchdahl inequality $2G M/c^2 R \leq 8/9$,
where $M$ is the total mass and $R$ the surface radius, which implies
that no star can have a larger gravitational
redshift than $2$ at the surface.
All relativistic stellar models satisfy the Buchdahl inequality, however,
the equality $2G M/c^2 R =8/9$, is associated with a
solution of constant mass-energy density and infinite central pressure.
Among other issues relying on differentiability of $\rho(p)$ in the limit $p\rightarrow \infty$ there is,
e.g., the mass-radius relationships for
large central pressures which we will discuss in the present paper.
The derivation of the Buchdahl inequality exemplifies that sometimes certain
rather unphysical solutions, associated with somewhat unphysical
equations of state, are important for physical problems.
Hence it is useful, indeed essential for
certain problems, to understand the entire solution space
associated with Eqs.~(\ref{oppvol}), including especially the limiting
solutions with infinitely small or infinitely large pressures.
In particular, in order to study generic behavior of solutions,
one has to understand the entire solution spaces of
large classes of equations of state.
As indicated, the TOV equation is rather unsuitable for this task.

The purpose of this paper is to introduce a formulation that avoids the
above problems and allows us to
(i) obtain a global picture of the entire solution space of a large set
of equations of state and to (ii) probe relationships between the equation
of state and global features like mass-radius properties.
To this end we will develop a framework based on the theory of dynamical systems; Eqs.~(\ref{oppvol})
will be reformulated
as a regular autonomous system of differential equations.
A formulation that allows such a global treatment of solutions
cannot exist for arbitrary barotropic equations of state.
However, the large classes of equation of state that can be covered
are of fundamental importance, as outlined in the following.

Let us start by making some dimensional
comparisons between General Relativity (GR) and Newtonian gravity.
The dimensions relevant for self-gravitating perfect fluid models
are mass, length, and time. In Newtonian theory and in GR
these are linked through the gravitational constant
$G$. Thus one can choose units so that one unit, e.g., mass, is expressed
in terms of the other two. In relativity yet another fundamental constant
appears -- the speed of light $c$. This constant reduces the number of
independent units to one, e.g., length. This restriction on the number
of independent scales is of great significance since scaling laws are of
great importance in gravitational theory, as in all branches of physics
(see, e.g., \cite{wies01} for references).

The Newtonian limit of~(\ref{potential}) and~(\ref{oppvol})
is obtained by defining the Newtonian potential%
\footnote{If one uses geometrized units $c=1$, $G=1$,
  there is no need for a distinction between $\Phi$ and $\Phi_N$.}
$\Phi_N := c^2\Phi$ and letting $c^2 \rightarrow \infty$, thus
setting the relativistic "correction" terms to zero. Note that
$\rho$ stands for the rest-mass density in the Newtonian case.
Requiring that these equations be invariant under scalings of time
and space forces a barotropic equation of state to be a polytrope
$\rho = K p^{n/(n+1)}$, where $K>0$ and $n\geq 0$ is the so-called
polytropic index. The appearance of $c$ in relativity reduce the
number of scale degrees of freedom from two to one since extra
terms arise in the equations as compared to Newtonian theory;
these terms originate from spacetime curvature and from pressure,
which now also contributes source terms, and are proportional to
$c^{-2}$. Hence polytropes do not leave the relativistic equations
invariant under scale transformations. Only the linear and
homogeneous equation of state $p = K\rho$ does.

Invariance implies symmetry, and symmetry implies mathematical simplification.
The existence of scale symmetries allows one to choose all variables except
one to be scale-invariant. Doing so leads to that the equation for the single
non-scale-invariant variable decouples and thus one obtains a reduced set of equations
for the scale-invariant variables. Hence polytropes and linear homogeneous equations
of state play a special role
in Newtonian theory and relativity, respectively ---
they are the mathematically simplest
perfect fluid models these theories admit, and the associated
equations can be formulated as symmetry-reduced problems.%
\footnote{Other equations of state may lead to greater
  simplifications for problems with extra symmetries, but it is only the polytropic and the linear
  equations of state that admit symmetry reductions in the general case in Newtonian gravity
  and relativity, respectively.}

The central role these equations of state occupy
is not due to gravity alone, but is also based on microscopic considerations
of matter models. Again scaling laws can be used to motivate their
special status, from a fundamental as well as phenomenological perspective.
Consider, e.g., Chandrasekhar's equation of state for white dwarfs,
where degenerate electrons contribute the pressure while the baryons supply the
mass. In this case there are two limits --- the non-relativistic electron limit,
which leads to a polytropic equation of state
with index $n=3/2$, and the relativistic electron limit,
which yields a polytrope with index $n=3$.
Another example is given by an ideal Fermi gas.
For extreme relativistic fermions $p = \rho c^2/3$ describes
the asymptotic behavior of the equation of state,
where the fermions supply
both the pressure and the mass-energy.

This discussion suggests that it is natural
to consider equations of state that are at least asymptotically
scale-invariant, i.e., asymptotically polytropic or linear, and
to classify equations of state according to their asymptotic limits;
$p\rightarrow 0$ and $p\rightarrow \infty$.
Their special status indicates that
asymptotically polytropic/linear equations of state
are the mathematically ``simplest'' equations of state
--- apart from their exact counterparts.%
\footnote{Again we are referring to generic models, special
  cases, e.g., special static spherically symmetric models, might admit
  "hidden" symmetries, and may thus be mathematically even simpler
  or lead to explicit solutions, see \cite{uggjanros95}, \cite{dellak98}
  and references therein.}

In GR, asymptotically linear equations of state possess asymptotic
invariance properties, however, surprisingly at first glance, also
asymptotic polytropes exhibit similar mathematical simplicity
in the low pressure regime.%
\footnote{This is satisfactory since in this regime asymptotically
polytropic equations of state are physically much more interesting.}
The reason for this is the
following: imposing spherical symmetry and requiring a static
perfect fluid equilibrium configuration results in
a close connection between
Newtonian gravity and GR. The spacetime symmetries prevent the existence
of gravitational waves and gravitomagnetic effects, thereby
reducing the number of degrees of freedom in GR to the same number as in
the Newtonian case. Asymptotically, the relativistic equations of
hydrostatic equilibrium coincide with their Newtonian counterparts provided that
$p/\rho c^2 \rightarrow 0$ when $p \rightarrow 0$, and hence it follows that
variables that are asymptotically scale-invariant in the Newtonian case
are also asymptotically scale-invariant in GR.%
\footnote{This idea was first introduced by
  Nilsson and Uggla \cite{nilugg00}, however, it will be exploited
further here.}
Accordingly, asymptotically polytropic equations of state
can be naturally included in the relativistic formalism.

The paper is structured as follows.
In Sec.~\ref{dynamicalsystemsformulation}
the dynamical systems approach to relativistic stellar models
is presented:
by introducing asymptotically scale-invariant variables
we reformulate the equations as a dynamical system.
In Sec.~\ref{equationsofstate} we discuss the assumptions
on the equations of state that are needed for the dynamical system
to become regular. Thereby the notion of asymptotically polytropic/linear
is made precise. The section concludes with a discussion about phase
transitions.
In Sec.~\ref{dynamicalsystemsanalysis} we investigate the state space
of the dynamical system and study its global properties.
The dynamical systems picture
is subsequently translated into a physical picture in Sec.~\ref{sec:trans}.
In Sec.~\ref{mass-radiustheorems},
we prove theorems concerning the relationship
between the equation of state and global mass and radius features.
To further illustrate our approach, we give several numerical examples
in Sec.~\ref{sec:ex}. Finally, in Sec.~8 we give a concluding discussion,
while we briefly compare the relativistic and Newtonian cases
in the Appendix.

\section{The dynamical system}
\label{dynamicalsystemsformulation}

Let us begin with some basic assumptions and definitions.
Throughout this paper we will assume that the perfect fluid is characterized
by a non-negative mass-energy density $\rho$ and a non-negative pressure $p$,
related by a continuous barotropic equation of state $\rho(p)$ that is
sufficiently smooth for $p>0$.%
\footnote{We will see in Sec.~\ref{equationsofstate} that, e.g.,
  ${\mathcal C}^2$ is sufficient, although
  this restriction can be weakened.
  Below we show how to handle even less restrictive situations
  like phase transitions.}
Define
\begin{equation}\label{indexdefs}
\Gamma_N(p) := \frac{\rho}{p}\, \frac{dp}{d\rho}
\:\,,\qquad\quad
\Upsilon(p) := \Gamma_N^{-1}(p) = \frac{p}{\rho}\, \frac{d\rho}{dp}
\:\,,\qquad\quad
\sigma(p) := \frac{p}{\rho c^2}\ ,
\end{equation}
where $\Gamma_N$ is the standard Newtonian adiabatic index%
\footnote{However, note that $\rho$ stands for the rest-mass density in
  the Newtonian case.}
, however, it turns out to be more convenient to use the inverse
index $\Upsilon$, which in addition naturally incorporates incompressible fluids
(for which $\Upsilon = 0$) into the formalism.
The frequently used polytropic index-function $n(p)$ is defined
via $\Upsilon(p) = n(p)/(n(p)+1)$, or equivalently
$\Gamma_N(p) = 1 + 1/n(p)$.

\begin{assumptions}
For simplicity, although only necessary asymptotically in our
framework, we assume that $d\rho/dp \geq 0$.
Then $\Upsilon \geq 0$ and $\rho(p)$ is
monotonic. If one impose the causality condition
$c_s \leq c$ on the speed of sound $c_s = \sqrt{dp/d\rho}$, then
$\sigma \leq \Upsilon$. Below, further assumptions will be imposed
by the dynamical systems formulation.
\end{assumptions}

A central idea of this paper is to obtain a regular
dimensionless autonomous system from Eqs.~(\ref{oppvol})
with a compact state space. To this end
we first "elevate" $r$ to a dependent variable and
introduce $\xi=\ln r$ ($r>0$) as a new independent variable.
We then make a variable transformation from $(m,p>0,r>0)$ to
the following three {\it dimensionless\/} variables,
\begin{equation}\label{uvom}
u = \frac{4\pi r^3 \rho}{m}\ , \quad
v = \frac{\rho c^2}{p}
\left(\frac{\frac{Gm}{c^2 r}}{1 - \frac{2Gm}{c^2 r}}\right)\ , \quad
\omega = \omega(p) \ .
\end{equation}
The new pressure variable $\omega(p)$ is
a dimensionless continuous function on $[0,\infty)$,
strictly monotonically increasing and sufficiently smooth
on $(0,\infty)$. We require that
$\omega(0)=0$ and $\omega \rightarrow \infty$ when
$p \rightarrow \infty$, and hence $\omega \geq 0$.
The remaining freedom will be used to adapt $\omega$ to
features exhibited by the classes of
equations of state of interest, cf.~Sec.~\ref{equationsofstate}.
The sign of $u$ and $v$ depends on whether $m$ is positive or negative.
In the following we restrict our attention to positive masses,
i.e., we investigate a perfect fluid solution only in that range of $r$
where it possesses a positive mass-function.%
\footnote{The region where $m(r)<0$ can be analyzed with the same
  dynamical systems methods that are going to be used in the following.
  The treatment turns even out to be considerably simpler, cf.~the
  case of Newtonian perfect fluids \cite{heiugg03}.}
Moreover, we require that $1 - 2 G m/(r c^2)>0$, i.e., we consider
only solutions to~(\ref{oppvol}) that give rise to a static
spacetime metric.%
\footnote{If a solution to~(\ref{oppvol}) satisfies $1 - 2 G m/(r c^2)>0$ initially at $r=r_0$
  for initial data $(m_0, p_0)$, then this condition holds everywhere.
  This has been proved (for regular solutions), e.g., in
  \cite{Rendall/Schmidt:1991}.
  Within the dynamical systems formulation this result
  can be established quite easily as we will see in
  Sec.~\ref{dynamicalsystemsanalysis}.
  Solutions violating the condition $1 - 2 G m/(r c^2)>0$ could be treated with
  the dynamical systems methods presented in this paper as well. However,
  we refrain from a discussion of such solutions here.}
Accordingly, we can assume $u>0$ and $v> 0$.

Starting from~(\ref{oppvol}) the transformation to the new variables $u,v,\omega$
yields the following system of equations:
\begin{subequations}\label{uvomeq}
\begin{align}
& \frac{du}{d\xi} = u\, (3 - u + \Upsilon\, h) \\
&\frac{dv}{d\xi} = v\,[-(1 - \Upsilon)h +
                       (u - 1)(1 + 2\sigma v) ] \\
&\frac{d\omega}{d\xi} = f \omega  h\:, \\
\label{hdef}
\mbox{where}\quad & h = h(u,v,\omega) = \frac{d \ln p}{d \ln r} =
-v(1 + \sigma)(1 + \sigma u)\: ,
\end{align}
\end{subequations}
and $f = f(\omega) = d\ln \omega/d\ln p\:|_{p(\omega)}$.
Also, $\sigma$ and $\Upsilon$ are understood as functions of $\omega$.

We now proceed by defining new bounded variables
$(U,V,\Omega) \in (0,1)^3$,
\begin{equation}\label{boundedvar}
U = \frac{u}{1+u}\ , \quad
V = \frac{v}{1+v}\ , \quad
\Omega = \frac{\omega}{1+\omega} \ .
\end{equation}

Introducing a new independent variable $\lambda$ according to
$d\lambda/d\xi = (1 - U)^{-1}(1 - V)^{-1}$ yields the dynamical system
\begin{subequations}\label{UVOmega}
\begin{align}
\label{Ueq}
&\frac{dU}{d\lambda} = U(1-U)[(1-V)(3-4U) - \Upsilon\,H] \\
\label{Veq}
&\frac{dV}{d\lambda} = V(1-V)[(2U-1)(1-V+2\sigma\,V) +
(1-\Upsilon)\,H] \\
\label{Omegaeq}
&\frac{d\Omega}{d\lambda} = -f\Omega(1-\Omega)H\, , \\[0.1cm]
\mbox{where}\quad
& H = H(U,V,\Omega) = (1 + \sigma)V(1 - U + \sigma U)\: ,
\end{align}
\end{subequations}
and where $\Upsilon$, $\sigma$, and $f$ are now functions depending on $\Omega$.

It turns out to be useful, indeed essential, to include the boundaries in our
analysis so that our compactified state space consists of the unit cube
$[0,1]^3$. To be able to discuss the dynamical system~(\ref{UVOmega})
in a straight forward manner
(e.g., to do a fixed point analysis) we require the
system to be ${\mathcal C}^1$-differentiable on $[0,1]^3$.
This is the case if $\Omega (1-\Omega) f(\Omega)$,
$\Upsilon(\Omega)$, and $\sigma(\Omega)$ are ${\mathcal C}^1$ on $[0,1]$.
In the next section we will show that a broad class of equations of state
satisfies these requirements.

\section{Equations of state}
\label{equationsofstate}

In this section the consequences of the ${\mathcal C}^1$-differentiability requirement are examined in detail.
There are three main building blocks:
firstly, the requirement that $\Omega (1-\Omega) f(\Omega)$ is ${\mathcal C}^1[0,1]$
is shown to imply that the choice of $\omega(p)$
in~(\ref{uvom}) is subject to certain restrictions;
secondly, the assumptions on $\Upsilon$ are formulated in a precise way:
$\Upsilon(\Omega)$ must be ${\mathcal C}^1$ in an admissible pressure
variable $\Omega$;
this assumption defines the classes of equations of state we can treat
in the dynamical systems framework;
thirdly, $\sigma$ is made ${\mathcal C}^1$ by the freedom
in choosing $\omega$.
The section is concluded by illustrative examples.

Part 1. The condition that $\Omega (1-\Omega) f(\Omega)$ is
${\mathcal C}^1[0,1]$ restricts the choice of the pressure variable $\omega(p)$ in~(\ref{uvom}).
This condition is satisfied
if and only if $f(\Omega)$ is ${\mathcal C}^0[0,1]$ and
${\mathcal C}^1(0,1)$, such that $\Omega df/d\Omega$ and $(1-\Omega) df/d\Omega$
vanish in the limit $\Omega\rightarrow 0$ and $\Omega\rightarrow 1$ respectively.
Expressed in terms of $f(p)$, a pressure variable
$\omega(p)$ is admissible, if the conditions $f(p)\in {\mathcal C}^0[0,\infty)$,
$\lim_{p\rightarrow\infty} f(p) < \infty$, $f(p)\in {\mathcal C}^1(0,\infty)$, and
$\lim_{p\rightarrow 0}d\ln f/d\ln p = \lim_{p\rightarrow\infty} d\ln f/d\ln p = 0$
are satisfied. It is not necessary, but convenient,
to choose $\omega$ in such a way that $f(\Omega)$
becomes strictly positive. Thus we require
$f_0 = f(0) >0$ and $f_1 = f(1) >0$ in the following.%
\footnote{Note that the Jacobian of the right hand side of~(\ref{UVOmega})
  contains $d/d\Omega\: [f(\Omega) \,\Omega (1-\Omega)]$,
  which when evaluated at $\Omega=0$ and $\Omega=1$
  equals $f_0$ and $f_1$, respectively. If these numbers are not zero
  the discussion of the dynamical system becomes easier.}

\begin{examples}
The simplest example of an admissible pressure variable is provided by
$\omega = (k p)^a$, where $a>0$ is a constant and where $k$ is a
positive constant carrying the dimension $[p]^{-1}$, so that $\omega$ becomes
dimensionless. In this example $f(p) = d\ln \omega/d\ln p \equiv  a$, which
obviously satisfies the imposed conditions. Other simple examples
include pressure variables described by $\omega(p) = (k p)^a (\pm \ln k p)^b$
for large $p$ or small $p$, respectively. Here either $a>0$ and $b\in\mathbb{R}$,
or $a=0$ and $b<0$.
\end{examples}

Part 2. Let us write the equation of state in the form
\begin{equation}\label{rhoexpl}
k \rho \,c^2 = \psi(k p)\:,
\end{equation}
where $k>0$ is a constant with dimension $[p]^{-1}$;
hence $\Upsilon(k p) = d\ln\psi(kp)/d\ln (kp)$. Recall that
we assume $d\rho/dp\geq 0$, i.e., $\Upsilon\geq 0$.

\begin{assumptions}
We assume that there exists a variable $\Omega$ (i.e., $\omega(p)$)
such that $\Upsilon(\Omega)$ is ${\mathcal C}^1[0,1]$. Moreover,
we require $\Upsilon_0 = \Upsilon(0) < 1$ and $\Upsilon_1 = \Upsilon(1) = 1$.
\end{assumptions}

\begin{remark}
The above assumptions restrict the equations of state to be
asymptotically polytropic for low pressures and asymptotically
linear at high pressures. Note that if one is only interested in
classes of perfect fluid solutions with bounded pressures, $p\leq
p_{{\rm max}}$, then restrictions at $\Omega=1$ are unnecessary,
since then $\Omega\leq \Omega_{{\rm max}} < 1$. The case
$\Upsilon_0=0$ corresponds to an equation of state that is
asymptotically incompressible when $p \rightarrow 0$. Above we
excluded the possibility $\Upsilon_0\geq 1$, which includes the
asymptotically linear case $\Upsilon_0=1$. This is not necessary,
but models with $\Upsilon_0\geq 1$ lead to solutions with infinite
masses and radii, and are therefore not particularly interesting
from an astrophysical point of view. Note, however, that
$\Upsilon(p)$ might be $\geq 1$ for some range of $p$
(corresponding to a negative polytropic index-function $n(p)$),
which happens, e.g., for equations of state that cover the
phenomenon of neutron drip.
\end{remark}

Let us now discuss the consequences of the assumptions for the equation
of state.
In a neighborhood of $\Omega=0$ let us write
$\Upsilon(\Omega) = \Upsilon_0 + \Omega \tilde{\upsilon}(\Omega)$,
or equivalently, $\Upsilon(\omega) = \Upsilon_0 + \omega \upsilon(\omega)$.
Then the above assumptions can be expressed as follows:
$\upsilon(\omega)$ is a continuous function, which is ${\mathcal C}^1$ away from $\omega=0$,
and satisfies $\lim_{\omega\rightarrow 0}\omega d\upsilon/d\omega = 0$.%
\footnote{Compare also with the discussion about admissible $\omega(p)$.}
By the definition of $\Upsilon$
we get $d\ln(p^{-\Upsilon_0} \rho)/d\ln p = \omega \upsilon$,
which, using $f=d\ln\omega/d\ln p$, gives rise to

\begin{equation}\label{eqofstatelow}
k \rho \, c^2  = s (k p)^{\Upsilon_0} \, \exp[\int f^{-1}(\omega) \upsilon(\omega) d\omega]
=: s (k p)^{\Upsilon_0} \,( 1 + g(\omega) )\:,
\end{equation}
where $s$ is a dimensionless constant.
The function $g(\omega)$ is ${\mathcal C}^1$, and even ${\mathcal C}^2$ for $\omega>0$, $g(0)=0$;
moreover, $\omega d^2g/d\omega^2$ vanishes as $\omega\rightarrow 0$. To first
order we have $g(\omega) = \int f^{-1}(\omega) \upsilon(\omega) d\omega$.

Analogously, in a neighborhood of $\Omega=1$, one can show that
the assumptions can be translated to
\begin{equation}\label{eqofstatehigh}
k \rho \, c^2  = (\gamma_1-1)^{-1} (k p) \,( 1 + \tilde{g}(\omega^{-1}) )\:,
\end{equation}
where $\gamma_1>1$ is a dimensionless constant, satisfying
$\gamma_1\leq 2$ if we require (asymptotic) causality.
Hence we have shown that the above assumptions on $\Upsilon$ are
satisfied if and only if there exists a pressure variable $\omega$
and functions $g$, $\tilde{g}$ with the above properties, such that
the equation of state~(\ref{rhoexpl}) can be written in the asymptotic
form~(\ref{eqofstatelow}) and~(\ref{eqofstatehigh}), respectively.

\begin{examples}
For a simple example consider the pressure variable $\omega = (k p)^a$ and
take $g(\omega) = const\, \omega$.
Then~(\ref{eqofstatelow}) reads $k \rho \, c^2  = (k p)^{\Upsilon_0} \,( 1 + const (k p)^a )$.
\end{examples}

\begin{remark}
If the assumptions hold, then
it is always possible to choose $\omega$ in such a way that $d\Upsilon/d\Omega$
vanishes as $\Omega\rightarrow 0$ and $\Omega\rightarrow 1$.
We can even achieve
\begin{equation}\label{Upsilonachievement}
\frac{d\Upsilon}{d\Omega} = O(\Omega^k) \quad (\Omega\rightarrow 0)
\qquad\mbox{and}\qquad
\frac{d\Upsilon}{d\Omega} = O((1-\Omega)^k) \quad (\Omega\rightarrow 1)
\end{equation}
for an arbitrary $k>0$.
This is based on the fact that if $\omega$ is an admissible pressure variable
then so is $\tilde{\omega}=\omega^l$ ($0<l<1$)%
\footnote{Its logarithmic derivative $\tilde{f}=d\ln\tilde{\omega}/dp$ satisfies $\tilde{f} = l f$.}
, and $d\Upsilon/d\tilde{\Omega} = O(\tilde{\Omega}^{(1-l)/l})$ for
$\tilde{\Omega}\rightarrow 0$,
and analogously for $\tilde{\Omega}\rightarrow 1$.
\end{remark}

Part 3. The third imposed condition is that $\sigma(\Omega)$
is ${\mathcal C}^1[0,1]$.
Firstly, the assumptions on $\Upsilon$ clearly imply that $\sigma(\Omega)$
is continuous with $\sigma_0 = \sigma(0) = 0$ and
$\sigma_1 = \sigma(1) = \gamma_1-1$.
Secondly, $\sigma(\Omega)$
is ${\mathcal C}^1[0,1]$, if $\Omega$ is chosen appropriately:
a straight forward computation of $d\sigma/d\Omega$ yields
$d\sigma/d\Omega = \sigma [\Omega (1-\Omega)]^{-1} f^{-1} (1-\Upsilon)$.
For $\Omega\rightarrow 1$ we obtain
$\lim_{\Omega\rightarrow 1}d\sigma/d\Omega = -(\gamma_1-1) f_1^{-1}
\lim_{\Omega\rightarrow 1} d\Upsilon/d(1-\Omega)$,
hence $\lim_{\Omega\rightarrow 1}d\sigma/d\Omega$ exists by the assumptions on $\Upsilon$.
In the limit $\Omega\rightarrow 0$ we obtain
$\lim_{\Omega\rightarrow 0}d\ln\sigma/d\ln\Omega = \lim_{\Omega\rightarrow 0}
(\Omega/\sigma)\, d\sigma/d\Omega= f_0^{-1} (1-\Upsilon_0)$.
If $f_0^{-1} (1-\Upsilon_0) > 1$, then $\lim_{\Omega\rightarrow 0} d\sigma/d\Omega$ exists
and $\lim_{\Omega\rightarrow 0} d\sigma/d\Omega =0$.
If $f_0^{-1} (1-\Upsilon_0) < 1$, then $d\sigma/d\Omega \rightarrow \infty$
for $\Omega\rightarrow 0$.%
\footnote{If $f_0 = (1-\Upsilon_0)$, then in many cases $\lim_{\Omega\rightarrow 0} d\sigma/d\Omega$ exists.
Note, however, that this does not hold in general.}
Therefore, provided that $\Omega$ is chosen so that
$f_0 < (1-\Upsilon_0) = 1/(1+n_0)$, we conclude that $\sigma(\Omega)\in {\mathcal C}^1[0,1]$.
In addition, in analogy with the discussion involving $\Upsilon$,
we can always achieve
\begin{equation}\label{sigmaachievement}
\frac{d\sigma}{d\Omega} = O(\Omega^k) \quad (\Omega\rightarrow 0)
\qquad\mbox{and}\qquad
\frac{d\sigma}{d\Omega} = O((1-\Omega)^k) \quad (\Omega\rightarrow 1)
\end{equation}
for arbitrarily large $k$.

\begin{remark}
Consider the equation of state~(\ref{rhoexpl}). We choose $\omega$
as a function of the dimensionless $k p$ as already indicated above, i.e.,
$\omega=\omega(k p)$.
Then the equation of state can be written in the implicit form
\begin{equation}\label{rhoimpl}
k p = \phi(\omega)\, ,\qquad\qquad k \rho c^2 = \chi(\omega) \:,
\end{equation}
where $\phi$ and $\chi$ may contain an arbitrary number of
dimensionless parameters. The derived quantities $f$, $\Upsilon$,
and $\sigma$ then read $f(\omega)=(d\ln\phi/d\ln\omega)^{-1}$,
$\Upsilon(\omega) = (d\ln\chi/d\ln\omega)
\,(d\ln\phi/d\ln\omega)^{-1}$, and $\sigma(\omega) =
\phi(\omega)/\chi(\omega)$. Note that $f(\omega)$,
$\Upsilon(\omega)$, $\sigma(\omega)$ are functions independent of
the dimensional parameter $k$; therefore the whole class of
equations of state~(\ref{rhoimpl}), or equivalently
(\ref{rhoexpl}), parameterized by $k$ is described by one
dynamical system~(\ref{UVOmega}) with specified $f(\Omega)$,
$\Upsilon(\Omega)$, and $\sigma(\Omega)$.
\end{remark}

\begin{example} Relativistic polytropes.
The relativistic generalization
of the Newtonian adiabatic index $\Gamma_N$ (as defined
in~(\ref{indexdefs})) is the adiabatic index
\begin{equation}
\Gamma(p) := \frac{1}{c^2}\:\frac{p+\rho c^2}{p}\:\frac{dp}{d\rho} = (1+\sigma) \,\Gamma_N(p)\:.
\end{equation}
Clearly, $\Gamma$ reduces to $\Gamma_N$ when $c\rightarrow\infty$.%
\footnote{Recall, however, that in the Newtonian case
          $\rho$ stands for the rest-mass density.}
Regarded as a function of $\Omega$, the adiabatic index $\Gamma$ becomes
a ${\mathcal C}^1[0,1]$ function for the considered class of
asymptotically polytropic/asymptotically linear equations of state.
In particular, $\Gamma(0) = \Gamma_N(0) = \Upsilon_0^{-1} = 1 +1/n_0$, and
$\Gamma(1) = 1+\sigma_1 = \gamma_1$.
Whereas $\Gamma_N=const$ defines the polytropic equations of state, $\Gamma=const$ gives
rise to the so-called relativistic polytropes,
usually written as
\begin{equation}\label{relpolold}
\rho c^2 = (p/K)^{1/\Gamma} c^2 + p/(\Gamma-1)\:,
\end{equation}
where $K$ is a positive parameter.
Requiring $d\rho/dp\geq 0$ and causality of the speed of sound $c_s=\sqrt{dp/d\rho}\leq c$
implies $1<\Gamma \leq 2$, where $\Gamma=2$ corresponds to the case
$c_s\rightarrow c$ when $p\rightarrow \infty$.
The class~(\ref{relpolold}) of equation of state can be expressed nicely in
the form~({\ref{rhoexpl}}), if we take $k = \left( c^2 (\Gamma-1) K^{-1/\Gamma}\right)^{-\Gamma/(\Gamma-1)}$:
\begin{equation}\label{relpol}
k\rho c^2 = \left(\frac{1}{\Gamma-1}\right) \:
\left((k p)^{1/\Gamma} + k p\right)\:.
\end{equation}
Obviously, the relativistic polytropes are of the
form~(\ref{eqofstatelow}) and~(\ref{eqofstatehigh}) by the simple
choice $\omega = (k p)^a$ with $a\leq (\Gamma-1)/\Gamma$, whereby
$\Upsilon(\Omega)$ and $\sigma(\Omega)$ are ${\mathcal C}^1[0,1]$.
To simplify the expressions for $\Upsilon(\Omega)$ and
$\sigma(\Omega)$ as much as possible we take $\omega = (k
p)^{(\Gamma-1)/\Gamma}$, i.e., $f=(\Gamma-1)/\Gamma$, which
results in $\sigma(\Omega) = (\Gamma-1)\Omega$ and
$\Upsilon(\Omega) = (1+(\Gamma-1) \Omega)/\Gamma$. Hence the right
hand side of the dynamical system~(\ref{UVOmega}) consists of pure
polynomials. Note that only one (dimensionless) parameter appears
in the dynamical system, namely $\Gamma$, i.e., the entire
one-parameter family parameterized by $k$ is represented by a
single dynamical system, as expected.
\end{example}

\begin{example} Ideal neutron gas.
The equation of state of a degenerate ideal neutron gas is implicitly given by
\begin{subequations}\label{neutron}
\begin{align}
\label{phineu}
k p &= \frac{1}{8\pi^2}\left( x(1+x^2)^{1/2}(\sfrac{2}{3} x^2 - 1) +
\ln[x+(1+x^2)^{1/2}]\right) \\
\label{chineu}
k \rho c^2 &= \frac{1}{8\pi^2}\left( x(1+x^2)^{1/2}(1+2x^2) -
\ln[x+(1+x^2)^{1/2}]\right)\; ,
\end{align}
\end{subequations}
where $x$ is essentially the Fermi momentum.%
\footnote{Namely, $x=p_F/m_n c$, where $m_n$ is the neutron mass.
  The constant $k$ is given by $k=\lambda_n^3/m_nc ^2$, where $\lambda_n= \hbar/m_n c$,
  for details see, e.g., \cite{Shapiro/Teukolsky:1983}, p.\ 23ff.}
A straight forward way to treat~(\ref{neutron}) is to choose $\omega =x$, whereby $\Omega=x/(1+x)$:
Firstly, $x$ is an admissible pressure variable, as $f(\Omega)$ is a sufficiently smooth function satisfying
$f =1/5 + O(\Omega^2)$ ($\Omega\rightarrow 0$) and $f = 1/4 + O((1-\Omega)^2)$ ($\Omega\rightarrow 1$).
Secondly, $\sigma(\Omega)$ is smooth with $\sigma = \Omega^2/5 + O(\Omega^4)$ and
$\sigma = 1/3 + O((1-\Omega)^2)$ as $\Omega\rightarrow 0,1$ respectively. In particular,
$\sigma_1=1/3$, i.e., $\gamma_1 = 4/3$.
Thirdly, $\Upsilon(\Omega)$ interpolates smoothly between $\Upsilon_0 =3/5$, i.e., $n_0=3/2$,
and $\Upsilon_1=1$. Moreover, $d\Upsilon/d\Omega = O(\Omega)$ and $d\Upsilon/d\Omega = O(1-\Omega)$
for $\Omega\rightarrow 0,1$.
Therefore, the ideal neutron gas is easily described
in the dynamical systems formalism, moreover, our
formalism also covers more sophisticated
equations of state that are used to model neutron stars.
\end{example}

To conclude this section we outline how
phase transitions and composite equations of state
can be incorporated into the dynamical systems framework.
Consider an equation of state $\rho=\rho(p)$ that is
piecewise continuous on $[0,\infty)$ and
piecewise sufficiently smooth on $(0,\infty)$.
This implies that at certain values $p_j$ of the pressure,
$\rho(p)$ makes a jump or a kink, corresponding
to a phase transition of first or second kind;
$\Upsilon(p)$ and $\sigma(p)$ exhibit similar behavior.
The associated ``weak solutions'' $p(r)$ and $m(r)$ of
equations~(\ref{oppvol})
are required to be continuous, while $\rho(r)$ is continuous only
if $\rho(p)$ is.
Also the variables $u, v$ and $U,V$ are continuous functions if
$\rho$ is; if $\rho$ jumps from some value $\rho_1$ to $\rho_2$ at
$p_j$, then $u,v$ make a jump according to
$u_1/u_2 = v_1/v_2 = \rho_1/\rho_2$, and analogously for $U,V$.
Choosing a conventional continuous pressure variable $\omega(p)$
we obtain piecewise continuous/smooth orbits, monotonic in $\Omega$,
that have possible jumps in $U,V$ at $\Omega_j = \Omega(p_j)$.

For practical purposes a slightly different viewpoint turns out to be
more convenient. Assume for simplicity that there is
only one jump of $\rho(p)$, at $p_j$. Choose two smooth equations
of state $\rho_{I}(p)$, $\rho_{II}(p)$, such that $\rho \equiv \rho_{I}$ for
$p < p_j$ and $\rho \equiv \rho_{II}$ for $p> p_j$.
Then the equation of state $\rho$ is simply obtained by
switching from $\rho_I$ to $\rho_{II}$ at $p_j$.
Correspondingly, we deal with two different state spaces,
in particular we can have two different pressure variables
$\Omega_I$ and $\Omega_{II}$.
>From this point of view, a fluid solutions associated with $\rho$
starts in one state space, and ends in the other state space;
the jump in between appears as a map from one state space to the other;
besides the jump in $U,V$ described above
it entails a jump in $\Omega$, from $\Omega_I(p_j)$ to $\Omega_{II}(p_j)$.
An explicit example will be given in Sec.~\ref{sec:ex}.

\section{Dynamical systems analysis}
\label{dynamicalsystemsanalysis}

In this section we study the dynamical system~(\ref{UVOmega});
in particular we investigate the global dynamics.
The arena for the analysis is the state space,
i.e., the unit cube $[0,1]^3$, which is endowed with different
vector fields and associated flows that depend on the equation of state.

\begin{figure}[ht]
\centering
\includegraphics[height=0.35\textwidth]{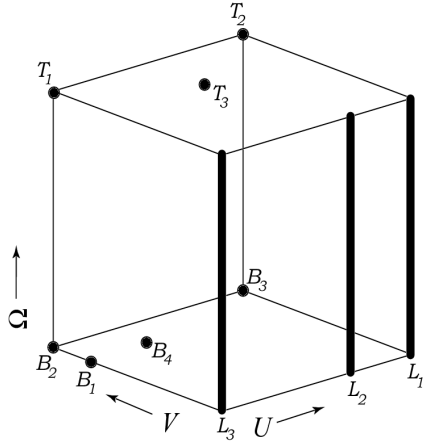}
\caption {The compact state space for the dynamical
system~(\ref{UVOmega}) with equilibrium points.
The position of $B_1$ and $B_4$ depends on $n_0$
($B_4$ is only present for $n_0>3$) and the position of
$T_3$ depends on $\gamma_1$.}
\label{fig:cube}
\end{figure}

Central to the dynamical systems analysis are the equilibrium points of the system.
The fixed points as well as their associated eigenvalues%
\footnote{The eigenvalues of the linearization of the system at the fixed point
  (together with the eigenvectors) characterize the flow of the dynamical system
  in a neighborhood of the fixed point (Hartman-Grobman theorem). For
  an introduction to the theory of dynamical systems,
  see e.g.~\cite{cra91}.}
are listed in Table~\ref{tab:UVcube}. Note that the equation of state
enters only via its asymptotic properties: both the
location of the fixed points and the eigenvalues depend only on $n_0$ and $\gamma_1$.
The state space and the fixed points are depicted
in Fig.~\ref{fig:cube}.

\begin{table}[ht]
  \begin{center}
    \begin{tabular}{|c|cccc|c|c|}
      \hline
      Fixed point & $U$ & $V$ & $\Omega$ &   & Eigenvalues & Restrictions \\  \hline
    & & & & &  & \\[-0.3cm]
      $L_1$ & 1 & 0 & $\Omega_0$ &   & $1\ , \ 1\ , \ 0$ & \\
      $L_2$ & $\frac{3}{4}$ & 0 & $\Omega_c$ &   &
      $-\frac{3}{4}\ , \ \frac{1}{2}\ , \ 0$ & \\
      $L_3$ & 0 & 0 & $\Omega_0$ &   & $3\ , \ -1\ , \ 0$ & \\
      $L_4$ & $U_0$ & 1 & 0 &  & $0 \ , \ -(1-U_0)\ , \ -f_0 (1-U_0)$ & $n_0 =0$ \\
      $B_1$ & 0 & $\frac{n_0+1}{n_0+2}$ & 0 &   &$-\frac{n_0 -3}{n_0+2}\ , \
      \frac{1}{n_0+2} \ , \ -\frac{f_0(n_0+1)}{n_0+2}$ &  \\
      $B_2$ & 0 & 1 & 0 &  &
      $-\frac{n_0}{n_0+1}\ , \ -\frac{1}{n_0+1}\ , \ -f_0$ & \\
      $B_3$ & 1 & 1 & 0 &  & $0\ , \ 0\ , \ 0$ & \\
      $B_4$ & $\frac{n_0 -3}{2(n_0 -2)}$ & $\frac{2n_0+2}{3n_0+1}$ & 0 &  &
      $\frac{\beta}{4}\left(5-n_0 \pm
        i \sqrt{b}\right)\ , \ -f_0\beta({n_0}^2-1)$
      & $n_0 > 3$\\
      $T_1$ & 0 & 1 & 1 &  & $-\gamma_1 \ , \ 2(\gamma_1-1)\ , \ f_1\gamma_1$ & $1<\gamma_1\leq2$\\
      $T_2$ & 1 & 1 & 1 &  & $\gamma_1(\gamma_1-1) \ , \ -2(\gamma_1-1)\ , \ f_1\gamma_1(\gamma_1-1)$ &$1<\gamma_1\leq2$ \\
      $T_3$ & $\frac{1}{2}$ & $\frac{2}{2+{\gamma_1}^2}$ & 1 &  &
            $-\frac{1}{{\gamma_1}^2+2}\frac{\gamma_1}{4}\left(3\gamma_1-2 \pm
        i \sqrt{c}\right)\ , \ \frac{1}{{\gamma_1}^2+2} f_1{\gamma_1}^2$
        &  $1<\gamma_1\leq2$ \\
        \hline
    \end{tabular}
  \end{center}
    \caption{Local properties of the fixed points.
    We use the abbreviations $b= -1-22n_0+7 n_0^2$,
    $\beta = \frac{n_0-1}{(n_0-2)(3n_0+1)}$, and
    $c =-{\gamma_1}^2+44\gamma_1-36$.}
    \label{tab:UVcube}
\end{table}

\begin{figure}[htp]
\centering
        \subfigure[$U=0$]{
        \label{U0side}
        \includegraphics[height=0.22\textwidth]{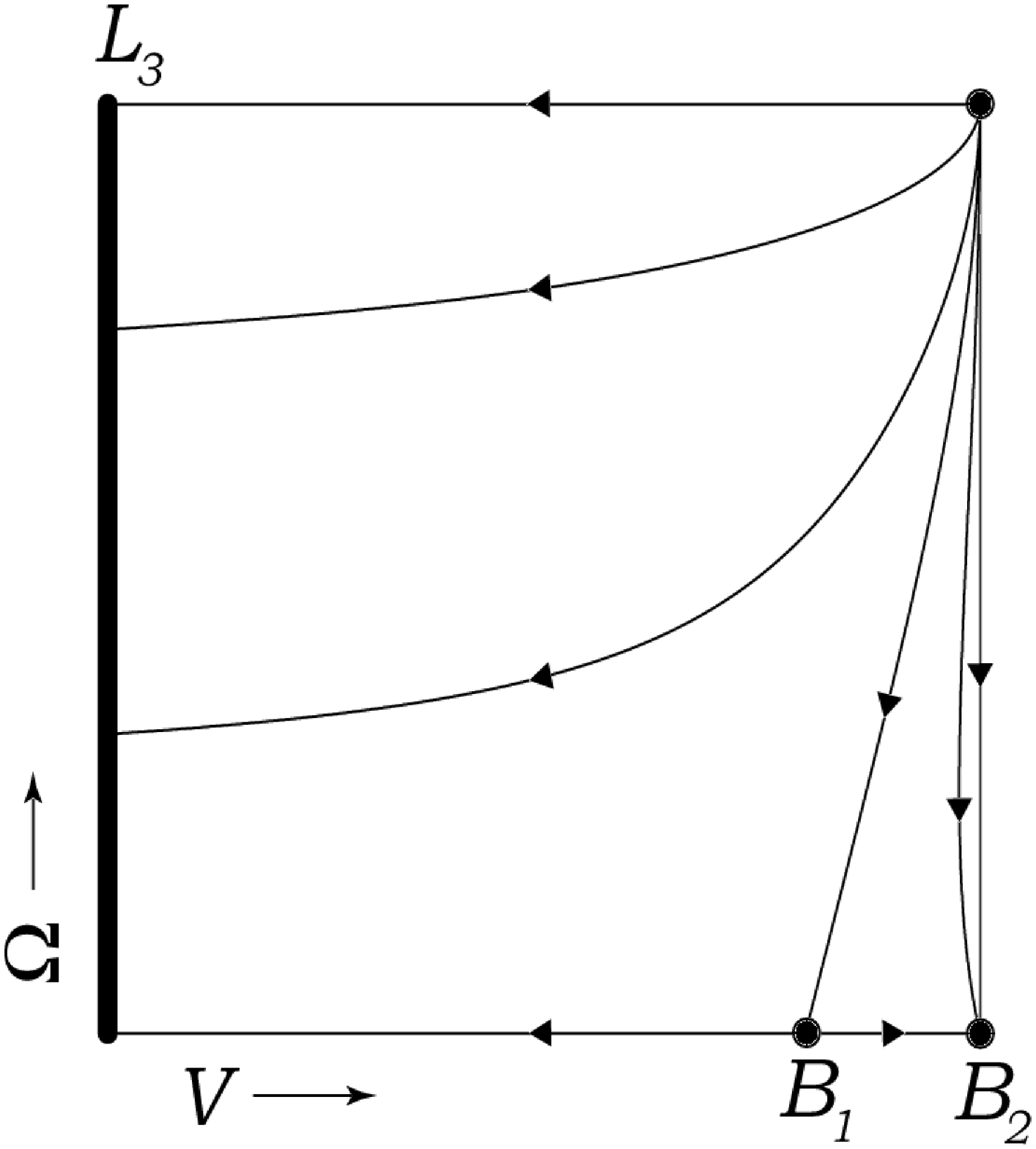}}\quad
        \subfigure[$U=1$]{
        \label{U1side}
        \includegraphics[height=0.22\textwidth]{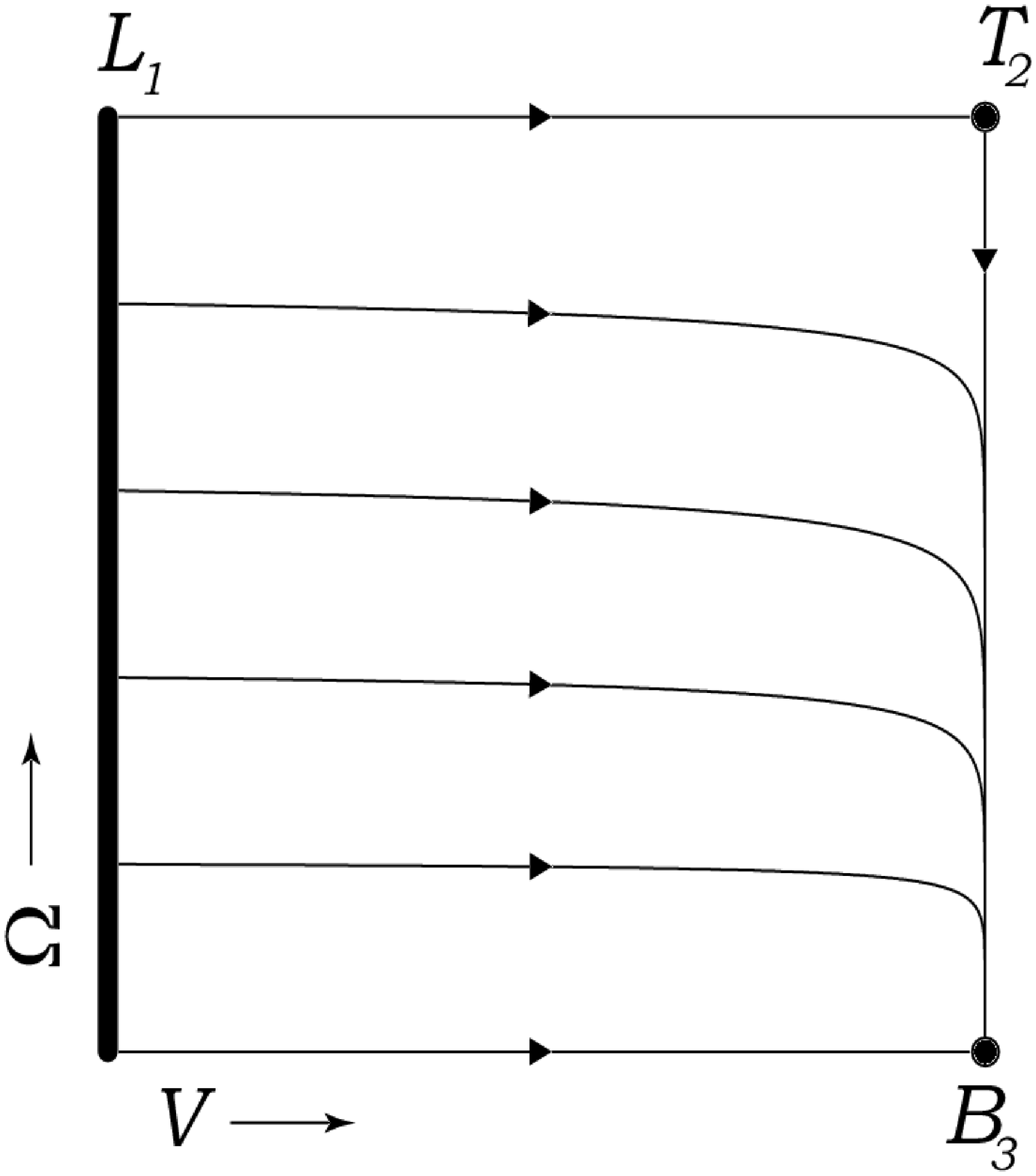}}\quad
        \subfigure[$V=0$]{
        \label{V0side}
        \includegraphics[height=0.22\textwidth]{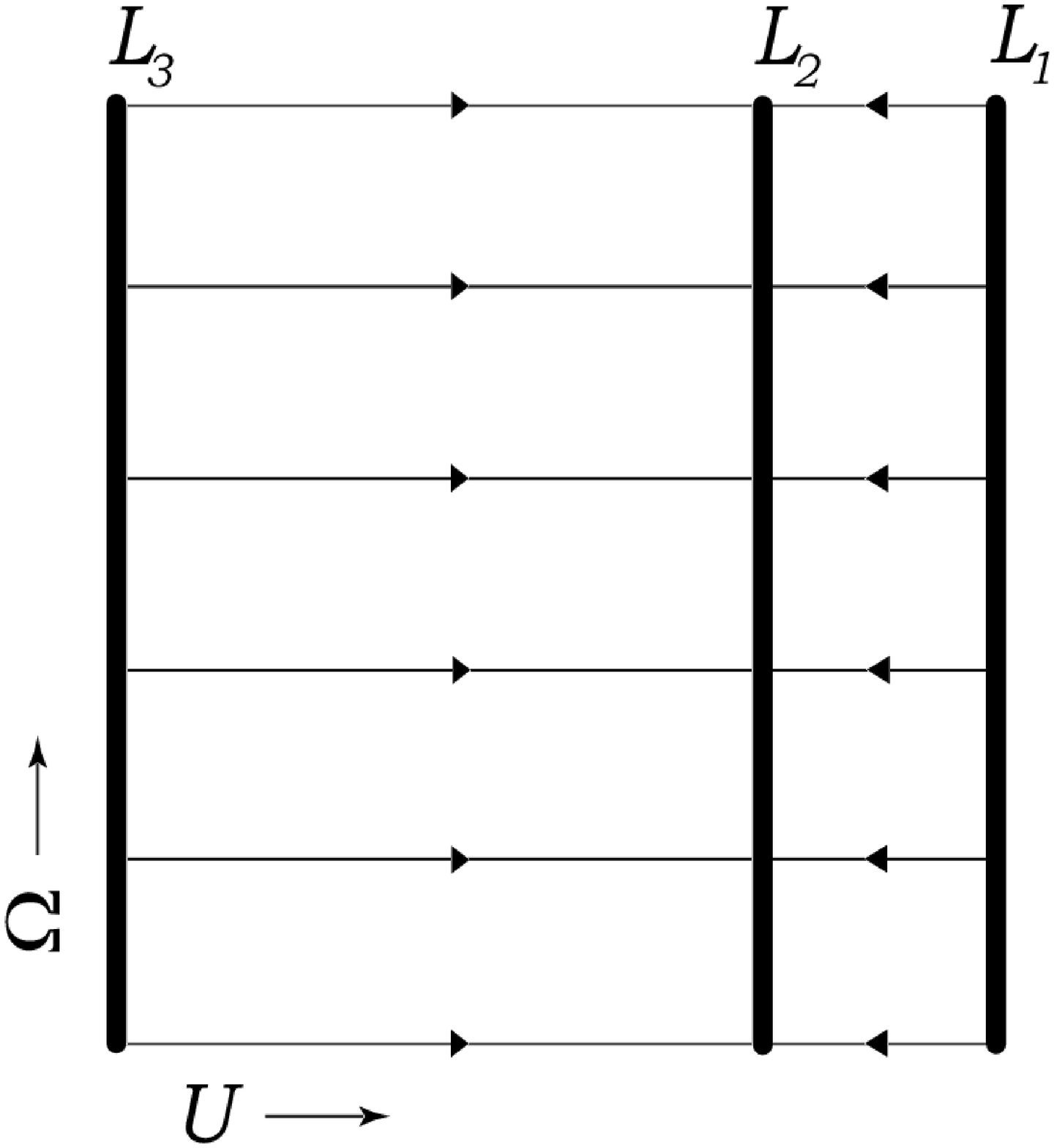}}\quad
        \subfigure[$V=1$]{
        \label{V1side}
        \includegraphics[height=0.22\textwidth]{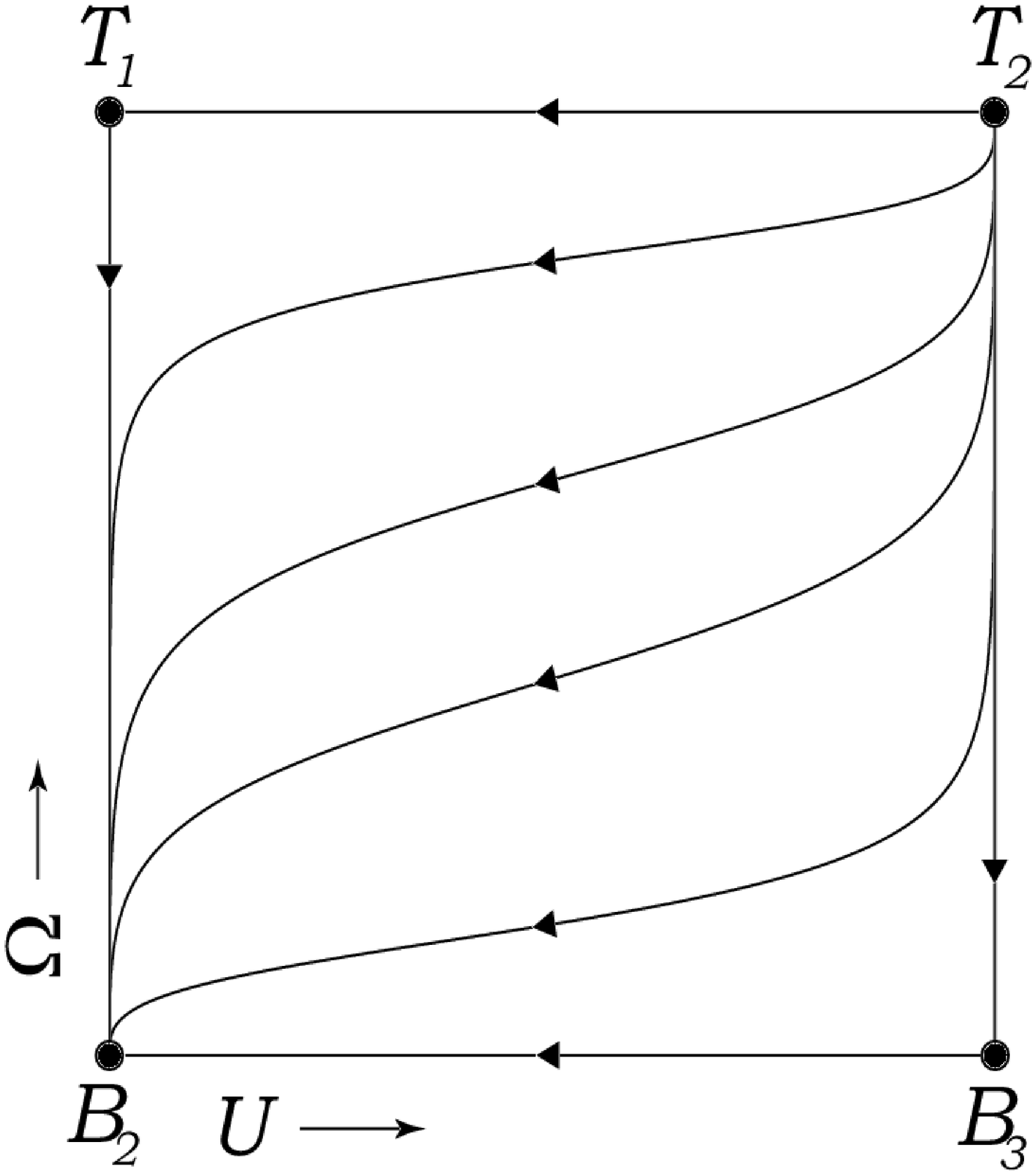}}\quad
        \subfigure[$n_0=0$]{
        \label{newtn0}
        \includegraphics[height=0.22\textwidth]{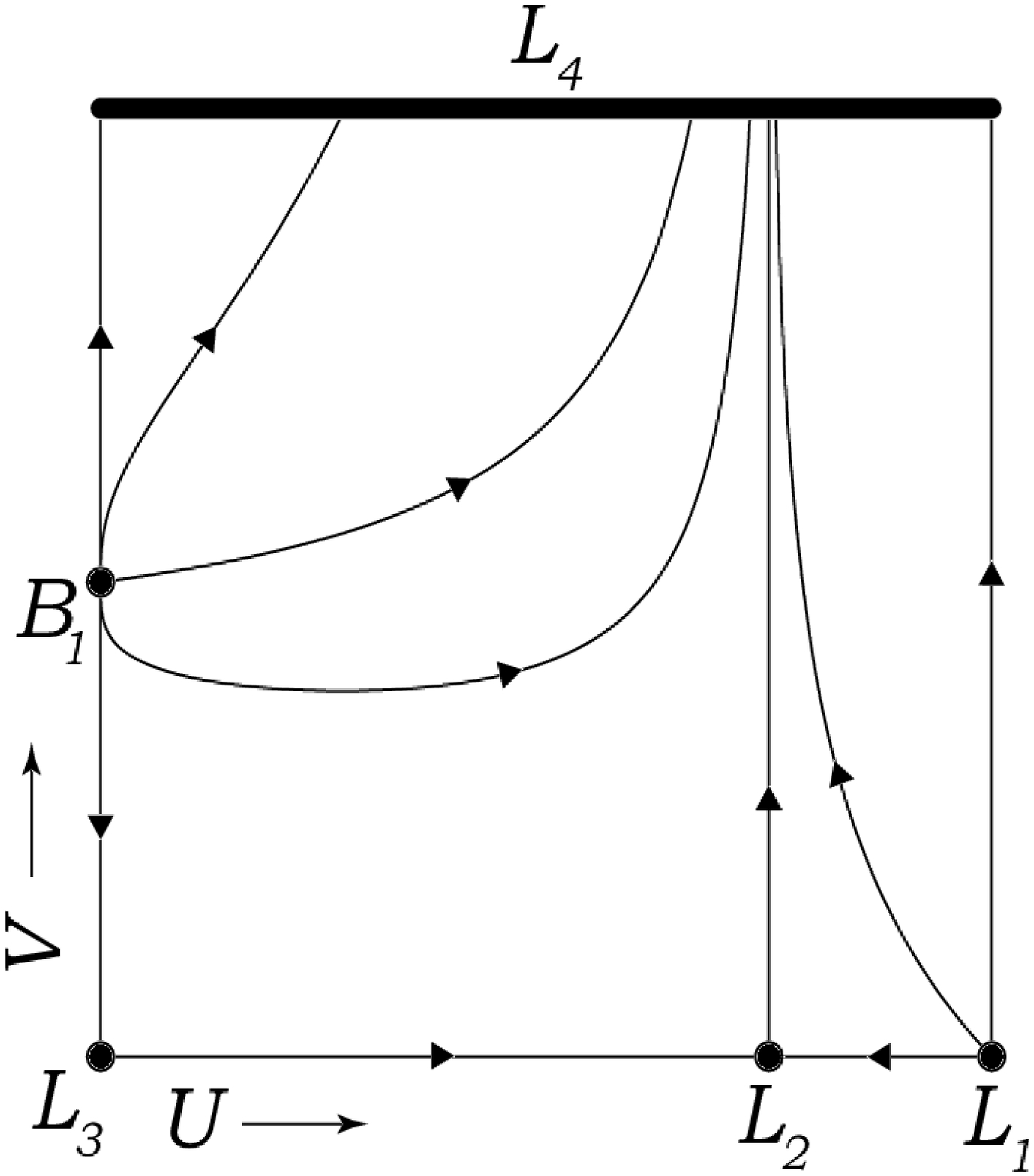}}\quad
        \subfigure[$0 <n_0 \leq 3$]{
        \label{newtn15}
        \includegraphics[height=0.22\textwidth]{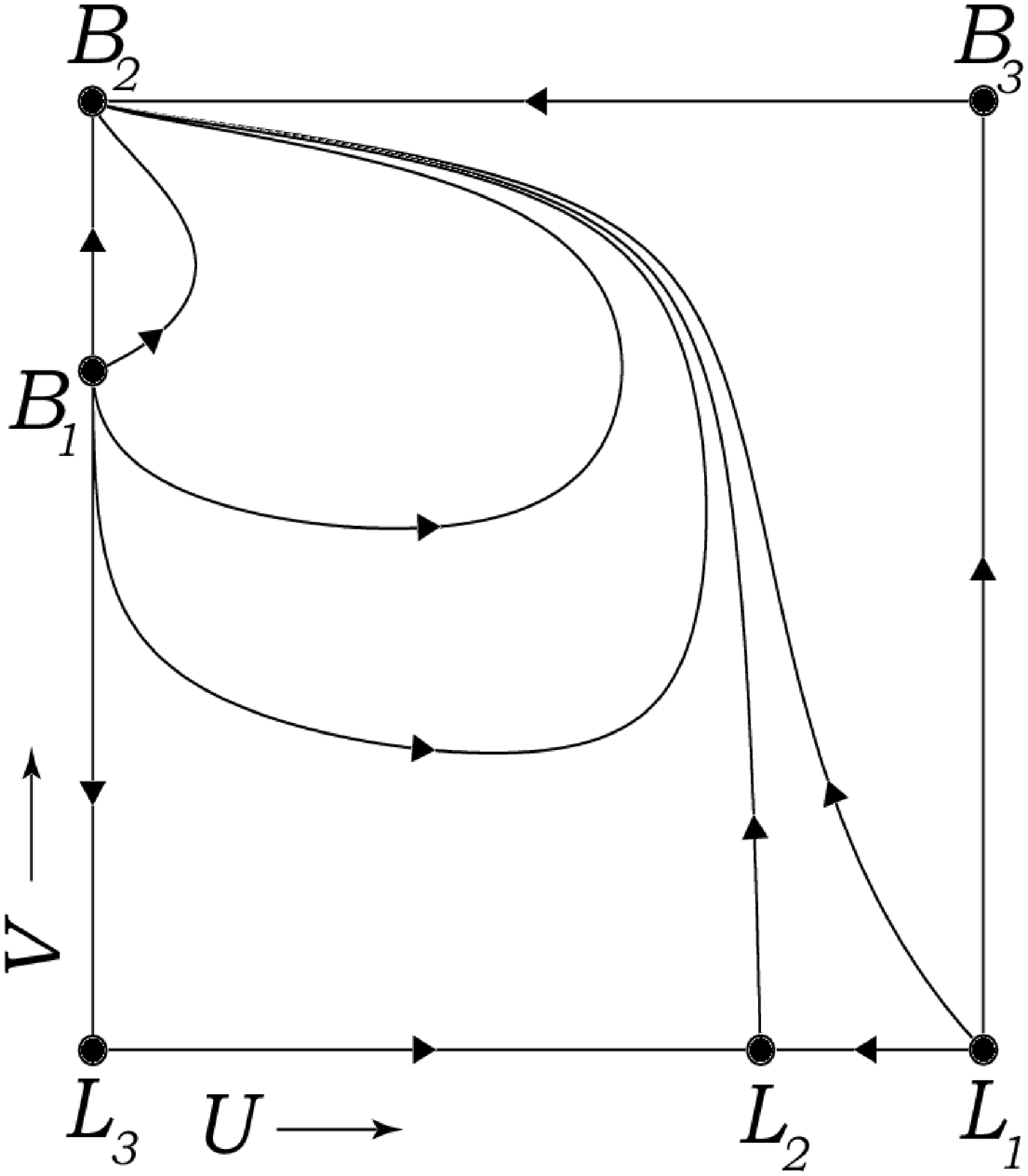}}\quad
        \subfigure[$3<n_0\leq(11+8\sqrt{2})/7$]{
        \label{newtn308}
        \includegraphics[height=0.22\textwidth]{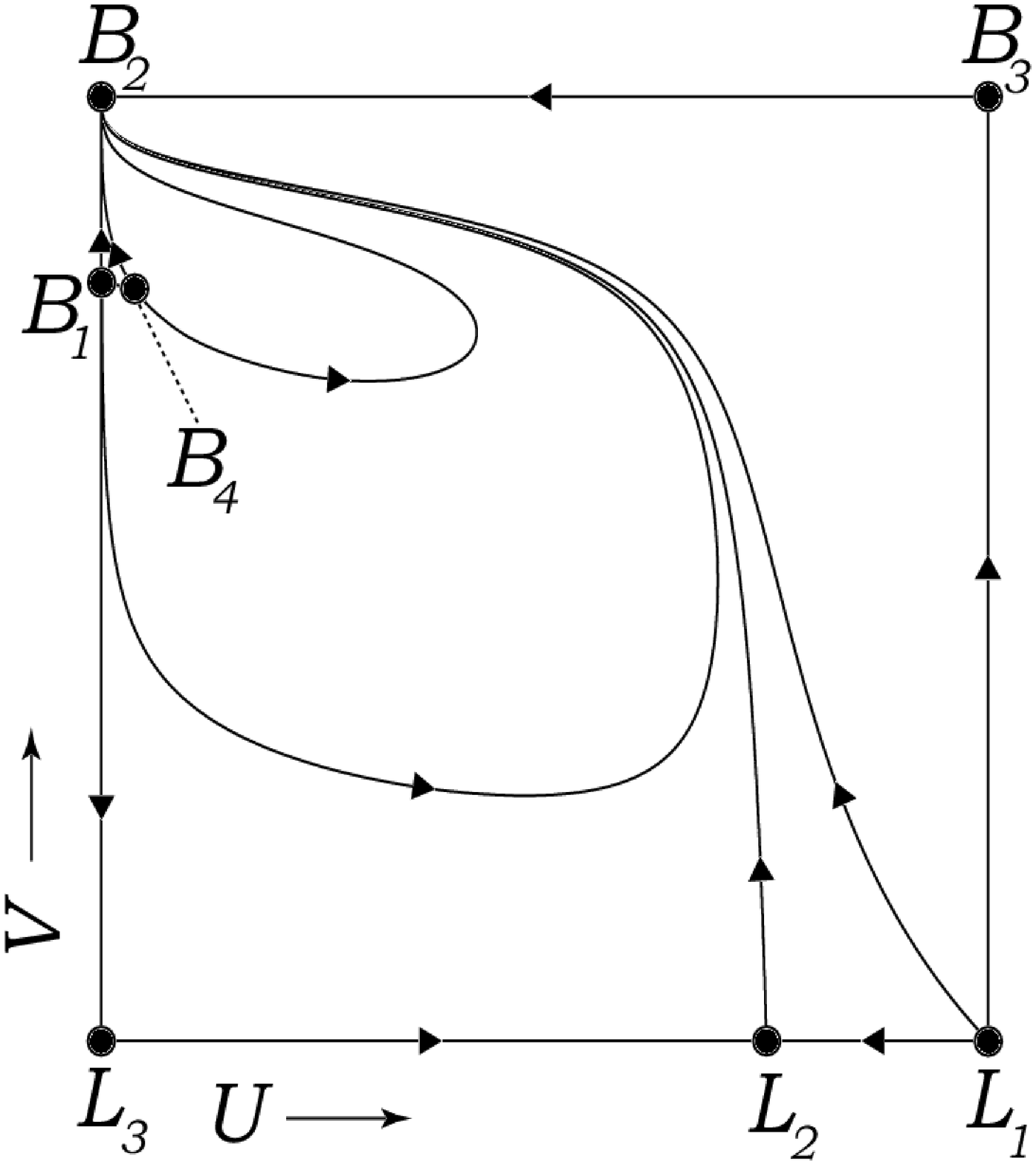}}\quad
        \subfigure[$(11+8\sqrt{2})/7<n_0\leq5$]{
        \label{newtn400}
        \includegraphics[height=0.22\textwidth]{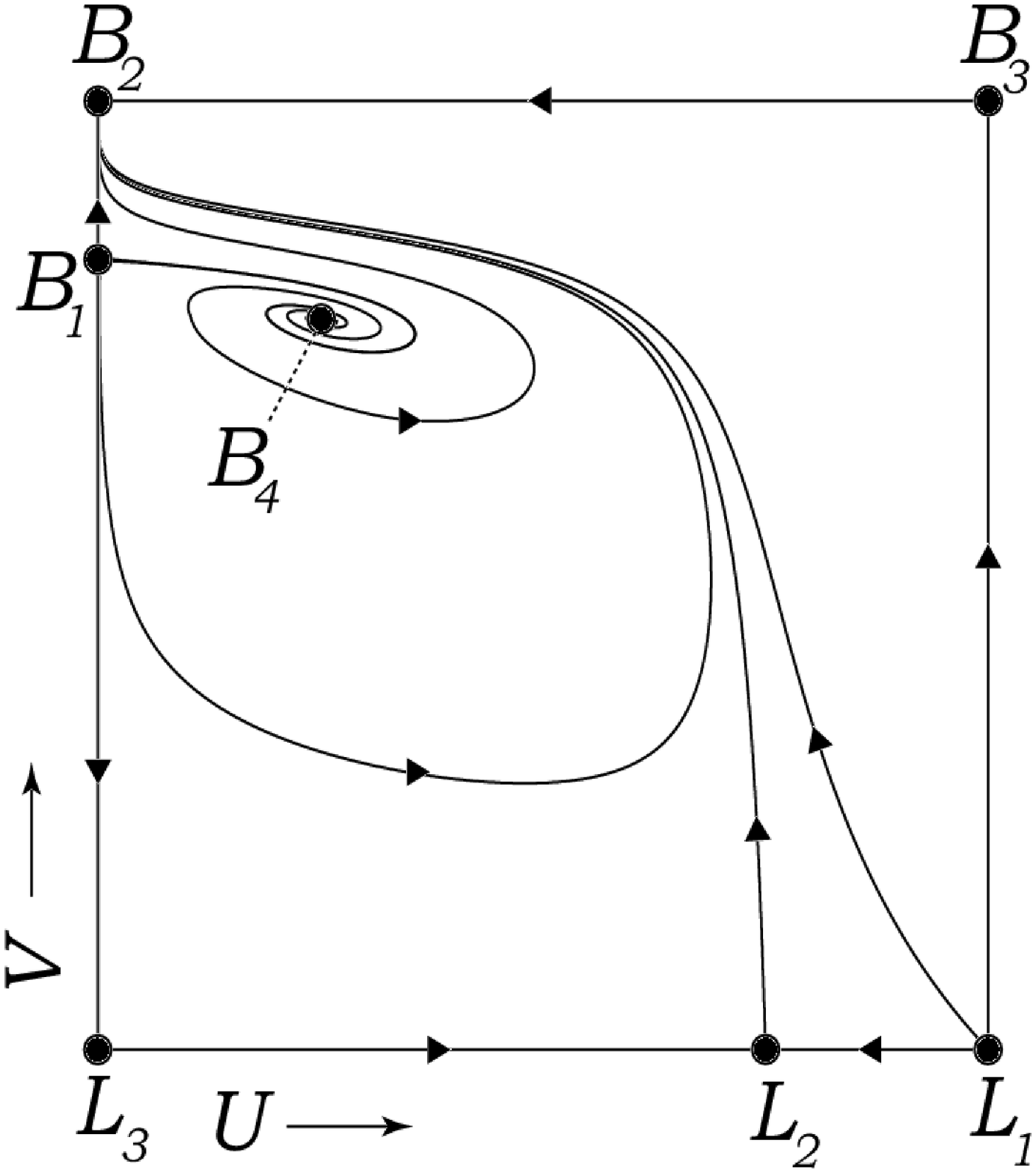}}\quad
        \subfigure[$n_0=5$]{
        \label{newtn500}
        \includegraphics[height=0.22\textwidth]{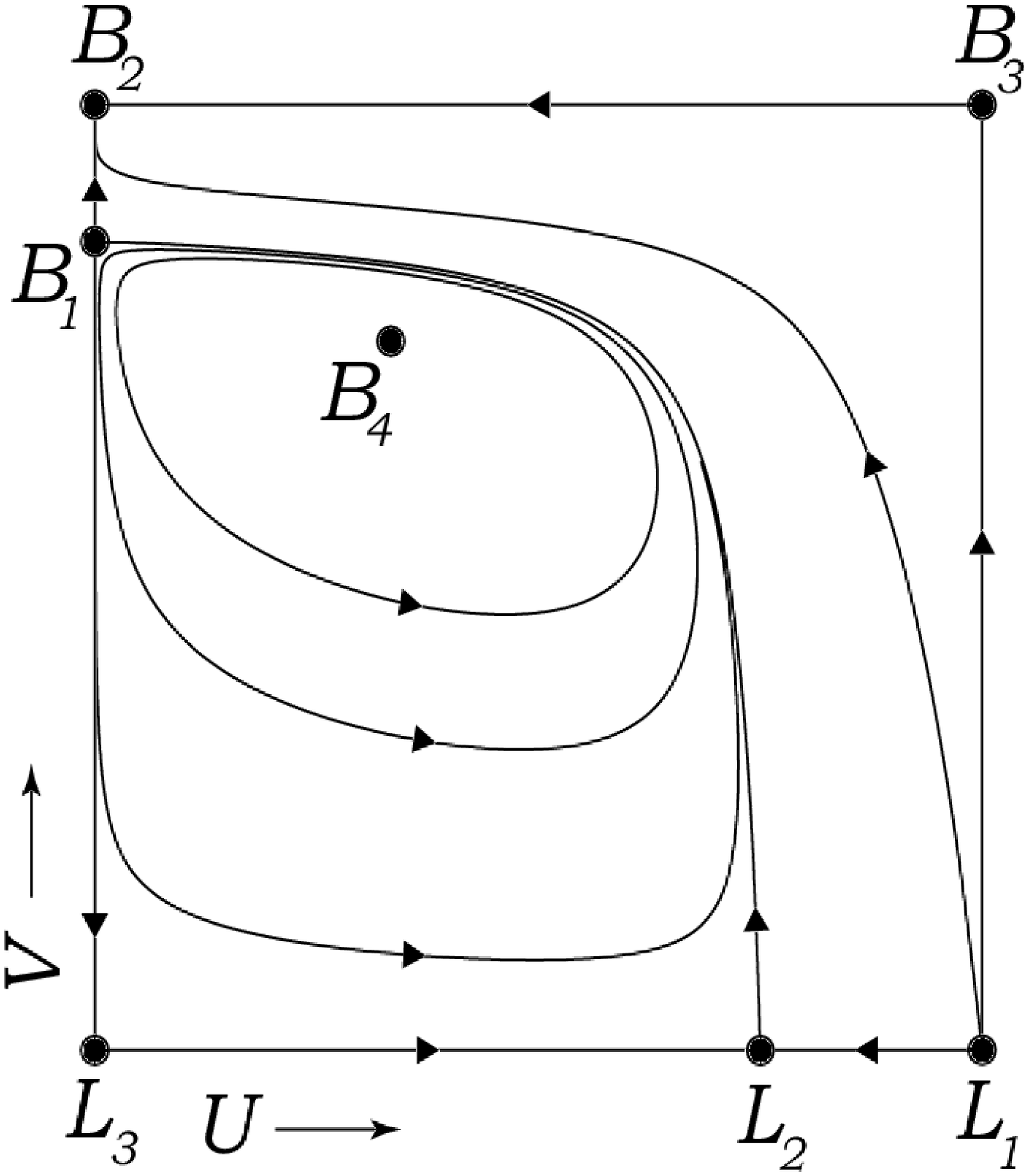}}\quad
        \subfigure[$n_0>5$]{
        \label{newtn800}
        \includegraphics[height=0.22\textwidth]{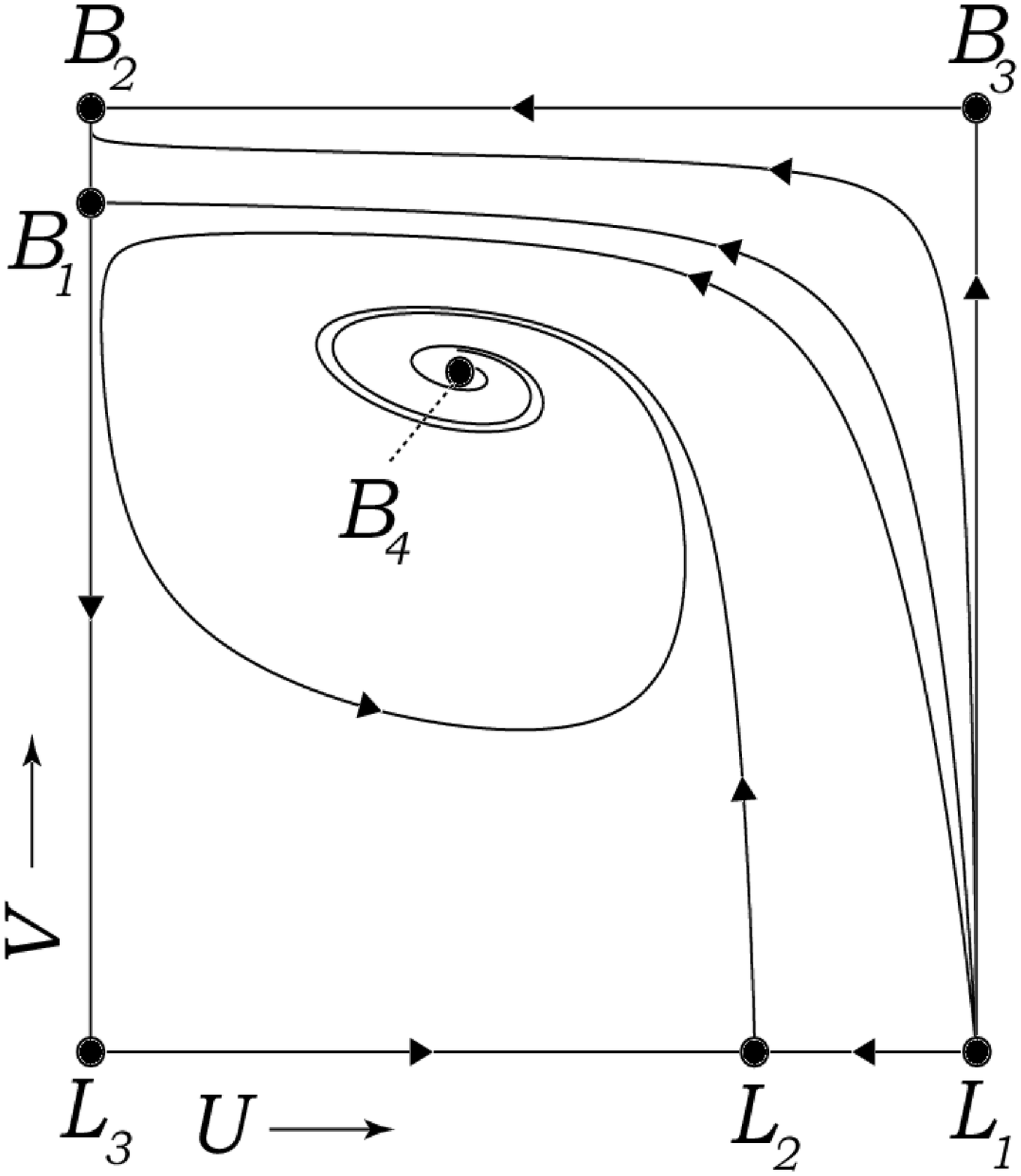}}\quad
        \subfigure[$\gamma_1=1.1$]{
        \label{Om1gamma11}
        \includegraphics[height=0.22\textwidth]{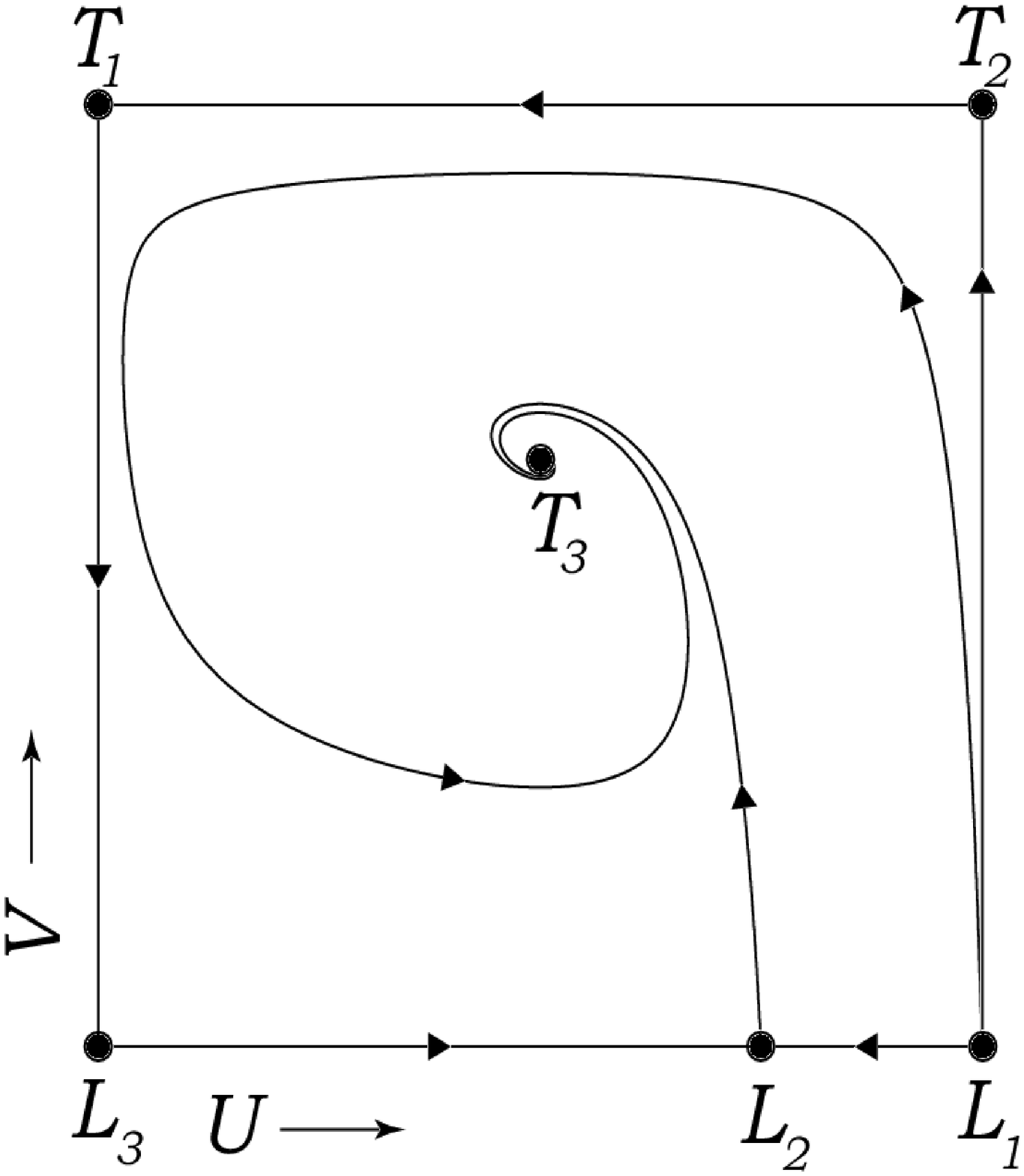}}\quad
        \subfigure[$\gamma_1=1.9$]{
        \label{Om1gamma19}
        \includegraphics[height=0.22\textwidth]{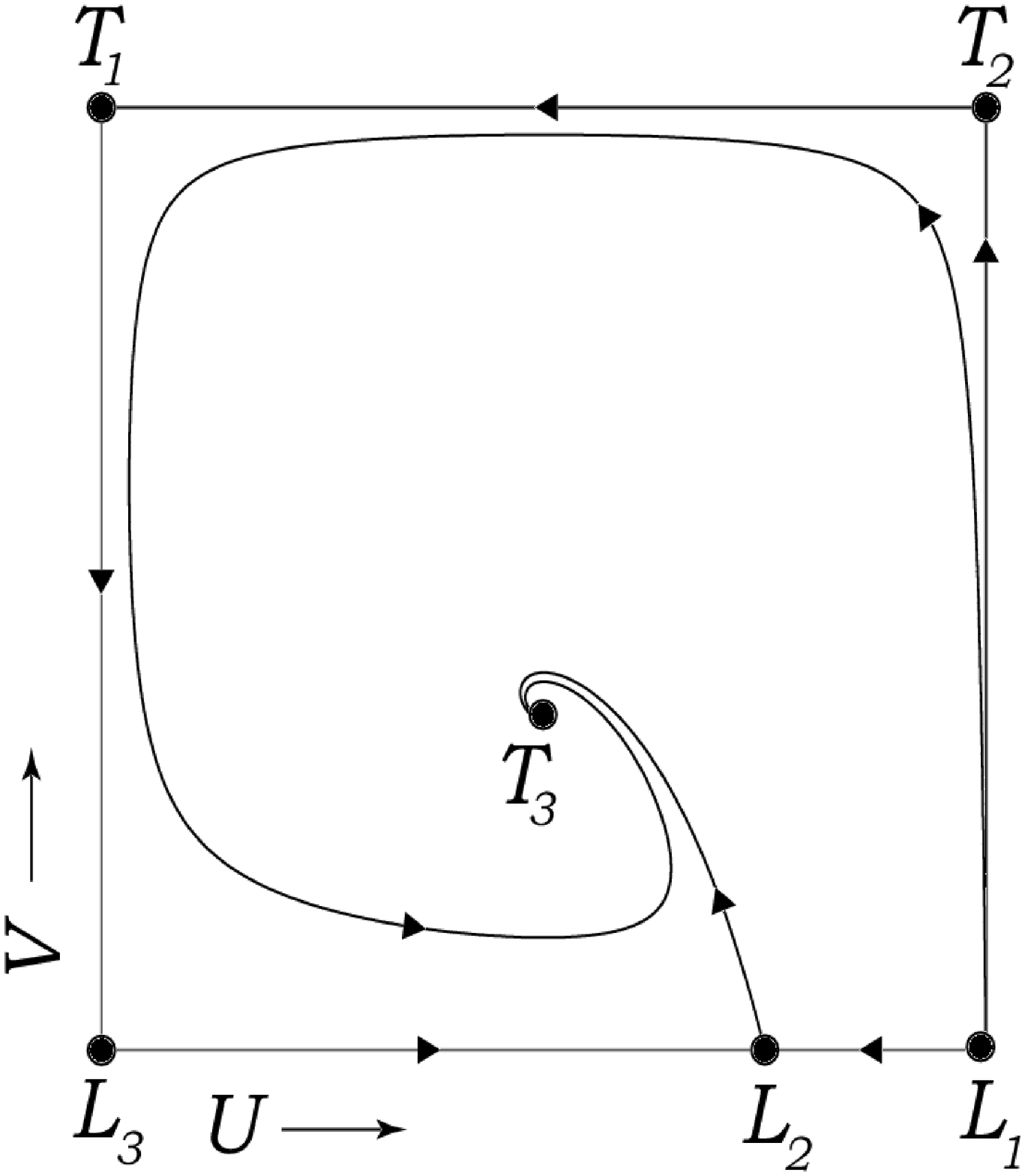}}\quad
        \caption{In (a), (b), (c), and (d) orbits for the four side faces are shown.
      The $U=0$, $U=1$, and $V=1$ faces involve $\sigma$ and   $\Upsilon$,
      which depend on the equation of state. For the plots the relativistic polytrope with $\Gamma=4/3$
      has been chosen; however, the qualitative structure is the same for all equations of state.
      In (e), (f), (g), (h), (i), and (j) orbits on the Newtonian subset $\Omega=0$ for different $n_0$
      are presented.
      The subfigures (k) and (l) show orbits on the $\Omega=1$ subset for some values of $\gamma_1$.}
        \label{fig:Cubefaces}
\end{figure}

We observe that all six faces of the cube are invariant
subspaces. On each of the four side faces, the induced dynamical system
possesses a simple structure;
the orbits on the side faces%
\footnote{Note that whereas the orbits in the interior of the cube
  represent perfect fluid solutions,
  the orbits on the side faces cannot be
  interpreted in physical terms. This is essentially
  because the variable transformation~(\ref{uvom}) cannot
  be inverted for, e.g., $u=0$ or $v=0$.
  In a modified dynamical systems formulation, where $\omega$
  is replaced by a variable depending on $\Phi$,
  certain side faces may be interpreted as representing vacuum solutions.}
are depicted in Fig.~\ref{fig:Cubefaces}.

The dynamical system on the $\Omega=0$ subset
is in fact Newtonian, cf.~Appendix~\ref{A}:
since $\sigma=0$ on $\Omega=0$,
$U$ and $V$ coincide with Newtonian homology variables
and thus the system represents the Newtonian
homology invariant equations
for an exact polytrope with index $n_0$.
On the $\Omega=0$ subset the flow of the dynamical system
exhibits rather complicated features.
In particular, $n_0$ serves as a bifurcation parameter:
there exist bifurcations for $n_0=0$, $n_0=3$, and $n_0=5$, as
indicated by the properties of the fixed points.
The case $n_0=5$ is exactly solvable: the orbits can be represented
implicitly by
\begin{equation}\label{n5int}
\Phi=\sqrt {U}{V}^{3/2}(U(8-9\,V)+7\,V-6)
-d (1-U)^{3/2}(1-V)^{5/2} = 0\:,
\end{equation}
where the parameter $d$ is required to satisfy $d \geq -9/8$.
The fixed point $B_3$ is represented by $d=-9/8$;
$-9/8< d <0$ defines a one-parameter family of closed
orbits, which we will henceforth denote by $C$;
$d=0$ characterizes the orbit that connects $L_2$ and $B_1$;
the remaining orbits correspond to
positive values of $d$, cf.~Fig.~\ref{newtn500}.
Typical orbits for a selection of values of $n_0$
are depicted in~Fig.~\ref{fig:Cubefaces};
cf.~also the analogous non-compact figures in
the standard literature, e.g., p.201 in~\cite{kipwei94}.

The dynamical system on the $\Omega=1$ subset can be regarded as describing
the relativistic scale-invariant equations for linear equations
of state $p = (\gamma_1-1)\rho c^2$,
a fact that reflects itself in the decoupling of the $(U,V)$-equations from the
system~(\ref{UVOmega}) for purely linear equations of state.
Typical orbits are depicted in Fig.~\ref{fig:Cubefaces}
(cf.~also, e.g.,~\cite{book:haretal65},
where an analogous non-compact phase portrait is given).

The {\it global dynamics} of the dynamical system~(\ref{UVOmega})
is determined by the fixed points on the boundaries of the state space,
which enables us to describe the asymptotics  of the interior orbits.
The statement is made precise in the following theorem.

\begin{theorem}\label{global}
The $\alpha$-limit%
\footnote{For a dynamical system $\dot{x} = f(x)$, the $\omega$-limit set
  ($\alpha$-limit) of a point $x$ is defined as the set of all accumulation
  points of the future (past) orbit of $x$. Limit sets thus characterize the
  asymptotic behavior of the dynamical system.}
of an interior orbit is a fixed point on $L_1$, $L_2$,
or the fixed point $T_3$.
The $\omega$-limit of an interior orbit is always located on $\Omega=0$,
and it is one of the fixed points $B_1$, $B_2$, or $B_4$, when $n_0\neq 5$. If
$n_0=5$, then the $\omega$-limit can also be
an element of the 1-parameter set of closed orbits $C$.
\end{theorem}

\proof
The proof of the theorem is based on the monotonicity principle
\cite{book:waiell97}:
if there exists a function $Z$ (defined on a closed, bounded, future (past) invariant
set $S$) that is monotonic along the flow of the dynamical system $\dot{x} = f(x)$,
then $\omega(S)$ ($\alpha(S)$) is contained in $\{x\in S | \dot{Z} = 0\}$.

The function $\Omega$ is such a monotone function on the
state space, i.e., $d\Omega/d\lambda <0$,
except on $\Omega=0$, $\Omega=1$, and $V=0$.
It follows that the $\omega$-limit of an interior orbit
is located on $\Omega=0$ or $V=0$,
while the $\alpha$-limit of an interior orbit must lie on $\Omega=1$
or $V=0$.
However, it follows from
the known orbit structure on $V=0$ and the local fixed point analysis
that $V=0$ cannot possess an $\omega$-limit.
Similarly, it follows that only $L_1$ and $L_2$
constitute $\alpha$-limits for interior orbits.

It remains to determine the $\omega$-limit ($\alpha$-limit) sets
for interior orbits on $\Omega=1$ and $\Omega=0$. Again we
make use of the monotonicity principle, however,
to simplify the discussion we use the uncompactified variables $(u,v)$.

On $\Omega=1$, a monotone function is given by%
\footnote{This monotone function was found through Hamiltonian methods, see~\cite{uggetal91}
  and chapter 10 in~\cite{book:waiell97}.}
\begin{equation}\label{monlin}
Z_3 = [-(\gamma_1-1) h_1(u,v) + 3\gamma_1 - 2]^2
(uv)^{-p_1}(1 + 2(\gamma_1-1)v)^{p_1-1} \ ,
\end{equation}
where the non-negative function $h_1$ and the non-negative constant $p_1$ are defined by
\begin{equation}\label{pg}
h_1(u,v) := \gamma_1(1+(\gamma_1-1)u)v = -h|_{\Omega=1}\; ,  \qquad
p_1 = \frac{4(\gamma_1-1)^2}{\gamma_1( 5\gamma_1 - 4)}\ .
\end{equation}
Its derivative along orbits reads
\begin{equation}\label{monlinder}
\frac{dZ_3}{d\xi} = -\frac{p_1}{2}\,(3\gamma_1-2)(2-h_1)^2
[(\gamma_1-1)h_1 + 3\gamma_1-2]^{-1}Z_3 \ .
\end{equation}
Application of the monotonicity principle yields that
only $T_3$ is a possible $\alpha$-limit set on $\Omega=1$.

We now turn our attention to the $\omega$-limit sets
of interior orbits, which we have shown to lie on $\Omega=0$.
For $n_0 \leq 3$ the function
$Z_1 = u v^3$ is monotone along orbits,
$dZ_1/d\xi = Z_1 (2 u + v(3-4 \Upsilon_0))$.
For $n_0>3$ define
\begin{equation}\label{Z2}
Z_2 = (uv)^{2/(n_0-1)}
\left[\frac{n_0-5}{(n_0-1)^2} +
\frac{2v}{(n_0+1)(n_0-1)} - \frac{v^2}{2(n_0+1)^2}-
\frac{uv}{(n_0+1)^2} \right] \ ,
\end{equation}
and observe that $dZ_2/d\xi = (n_0-5)\,(n_0+1)^{-2}(n_0-1)^{3}
((uv)^{2/(n_0-1)}\big(2(n_0+1)-(n_0-1)v\big)^2$,
i.e., $Z_2$ is monotonic on orbits for $3<n_0\neq 5$.
(Note that the flow is non-vanishing on $v=2(n_0+1)/(n_0-1)$ except for at $B_4$.)
Thus, by the monotonicity principle, taking also the solution structure on the boundaries
into account, only the fixed points $B_1$, $B_2$, $B_4$ can serve as
$\omega$-limit sets on $\Omega=0$.
The exceptional case is $n_0=5$:
then there also exists a 1-parameter family, $C$, of closed orbits (periodic solutions)
described by~(\ref{Z2}) (see also~(\ref{n5int})),
which acts as an $\omega$-limit set for interior solutions.
\proofend

\begin{theorem}\label{n03}
If $0<n_0\leq 3$ $(n_0=0)$, then all orbits end at $B_2$ $(L_4)$.
\end{theorem}

\proof
We distinguish two cases, $n_0<3$ and $n_0=3$.  For $n_0<3$ the
proof is trivial; each interior orbit
converges to $B_2$ for $0<n_0<3$ (to $L_4$ when $n_0=0$), which
follows from the proof of Theorem~\ref{global} and the local dynamical
systems analysis (Table~\ref{tab:UVcube}).

In the case $n_0=3$, $B_1$ is not hyperbolic. Nevertheless,
by applying center manifold theory, we show that
no interior orbit converges to $B_1$ as $\lambda\rightarrow\infty$:

It is advantageous to investigate the problem
in adapted uncompactified variables $\{x\geq 0,y,z\geq 0 \}$,
defined by
\begin{equation}\label{oldvars}
u = x (n_0-2)/(n_0+1) \qquad v = (n_0+1)-x+y \qquad \omega = z \:,
\end{equation}
The fixed point $B_1$ is represented by $(0,0,0)$ and the dynamical system~(\ref{uvomeq}) reads
\begin{equation}\label{sysinxyz}
\begin{array}{lcl}
\dot{x} = & & + \quad N_x(x,y,z) \\[3pt]
\dot{y} = & y & +\quad N_y(x,y,z)\\[3pt]
\dot{z} = & -f_0 (1+n_0) z & + \quad N_z(x,y,z) \:,
\end{array}
\end{equation}
where $N_x$, $N_y$, $N_z$ denote the nonlinear terms. $N_x$ is given by
\begin{equation}
N_x = x(3-4\Upsilon(z) - x/4 + (x-y) \Upsilon(z)) + C_1(x,y,z) \sigma(z) + C_2(x,y,z) \sigma^2(z)\:.
\end{equation}
A center manifold of the system~(\ref{sysinxyz}), i.e., an invariant manifold tangential
to the center subspace at the fixed point, is represented by
$\{(x,h_y(x),h_z(x))|0\leq x< \varepsilon \}$ satisfying
\begin{equation}
\partial_x h_y(x)\; \dot{x}(x,h_y(x),h_z(x)) = \dot{y}(x,h_y(x),h_z(x))
\qquad
\partial_x h_z(x)\; \dot{x}(x,h_y(x),h_z(x)) = \dot{z}(x,h_y(x),h_z(x)) \:,
\end{equation}
and the tangency conditions $h(0)=0$, $\partial_x h(0)=0$. We obtain
\begin{equation}
h_y(x) = -x^2/2 - x^3 + O(x^4) \qquad h_z(x) \equiv 0\:.
\end{equation}
Note in particular that the center manifold lies in the Newtonian subset of the state space.
The center manifold reduction theorem states that the flow of the full nonlinear
system is locally equivalent to the flow of the reduced system
(see, e.g., \cite{cra91})
\begin{equation}\label{redsysinxyz}
\begin{array}{lcl}
\dot{x} &= & N_x(x,h_y(x),h_z(x)) = x^2/2 + O(x^3) \\[3pt]
\dot{y} & =& y \\[3pt]
\dot{z} &= & -f_0 (1+n_0) z  \:.
\end{array}
\end{equation}
The flow given by~(\ref{redsysinxyz}) clearly prevents
interior solutions $(x>0, y, z>0)$
from converging to $B_1$, which proves the claim.
\proofend

Collecting the results of the local and global dynamical systems analysis
yields the following.
Interior orbits originate from $L_1$ (a two-parameter set);
$L_2$ (a one-parameter set); and $T_3$ (a single orbit). They
end at $B_1$ (when $n_0>3$; a one-parameter set); $B_2$ (a two-parameter set);
$L_4$ (when $n_0=0$; a two-parameter set);
$B_4$ (when $n_0>3$; a single orbit when $n_0\leq5$ and a two-parameter set
when $n_0>5$); and at the periodic curves $C$ (when $n_0=5$; a
two-parameter set).

An orbit in the state space corresponds to a perfect fluid
solution. Note that the perfect fluid solution is
only represented in that range of $r$ where $p>0$,
i.e., the fluid body, and where $m>0$;
cf.~Sec.~\ref{dynamicalsystemsformulation}.
If the configuration has a finite radius $R$, then
the fluid solution is joined to a vacuum solution,
a Schwarzschild solution, at $r=R$.

The one-parameter set of {\it regular} perfect fluid solutions, i.e.,
solutions of~(\ref{oppvol}) with a regular center of spherical symmetry,
appears in the state space as the one-parameter set of orbits
that originate from $L_2$.
When $d\rho/dp\geq 0$ is assumed, $\rho(r) \leq \bar{\rho}(r)$ holds
for regular solutions, where $\bar{\rho} = 3m/(4\pi r^3)$ is the ``average density''.
Translated to the state space variables we obtain $U\leq 3/4$.
In order to show this inequality within the dynamical systems framework,
consider the surface $U=3/4$ and the flow through this surface:
$dU/d\lambda|_{U=3/4} = -3\Upsilon(1+\sigma)(1+3\sigma)V/64\leq 0$,
since $\Upsilon\geq 0$ when $d\rho/dp\geq 0$. Hence, $U=3/4$ acts as a
``semi-permeable membrane'' for the flow of the dynamical system.
By noting that the unstable subspaces at $L_4$ are characterized by $U\leq 3/4$
the claim is established.%
\footnote{We observe that in the incompressible
  case, $\rho=const$ (i.e., $\Upsilon =0$), the regular solutions are located on $U=3/4$. The equations
  are explicitly solvable in this case. Note, however, that we have to cut the
  state space at some value of $\Omega$, because $\Upsilon=0$ is not
  compatible with the requirement $\Upsilon_1=1$ to obtain a ${\mathcal C}^1$
  dynamical system up to and including $\Omega=1$.}

If a perfect fluid solution satisfies $1 - 2 G m/(r c^2)>0$ initially at $r=r_0$
for initial data $(m_0, p_0)$, then this condition holds everywhere.
For a proof (involving regular solutions) see,
e.g.,~\cite{Rendall/Schmidt:1991}. In the state space picture this
result can be established quite easily. By construction,
the initial data is a point in the interior of the state space
and the orbit passing through this point represents the associated
perfect fluid solution.
Since $V<1$ for the orbit, we obtain that
\begin{equation}
1 - \frac{2 G m}{r c^2} = \frac{1-V}{1-V+2 \sigma V} >0
\end{equation}
holds everywhere within the fluid body.
We can even show that $1-2 G m/(r c^2) \geq  const >0$ for all $r$:
the considered orbit can have three possible end points;
suppose that the orbit ends in the fixed point $B_2$, which is
the only non-trivial case since then $V\rightarrow 1$
when $\lambda\rightarrow\infty$.
We will see in the subsequent section that
such an orbit gives rise to a perfect fluid solution possessing
a finite radius $R$ and a total mass $M$,
which are related by $G M/(R c^2) = B/(2B+s)$, where $B$ and $s$ are
positive numbers. Hence the claim is established.

\begin{figure}[h]
\centering
        \subfigure[$\Gamma=3/4$]{
        \label{G43}
        \includegraphics[height=0.35\textwidth]{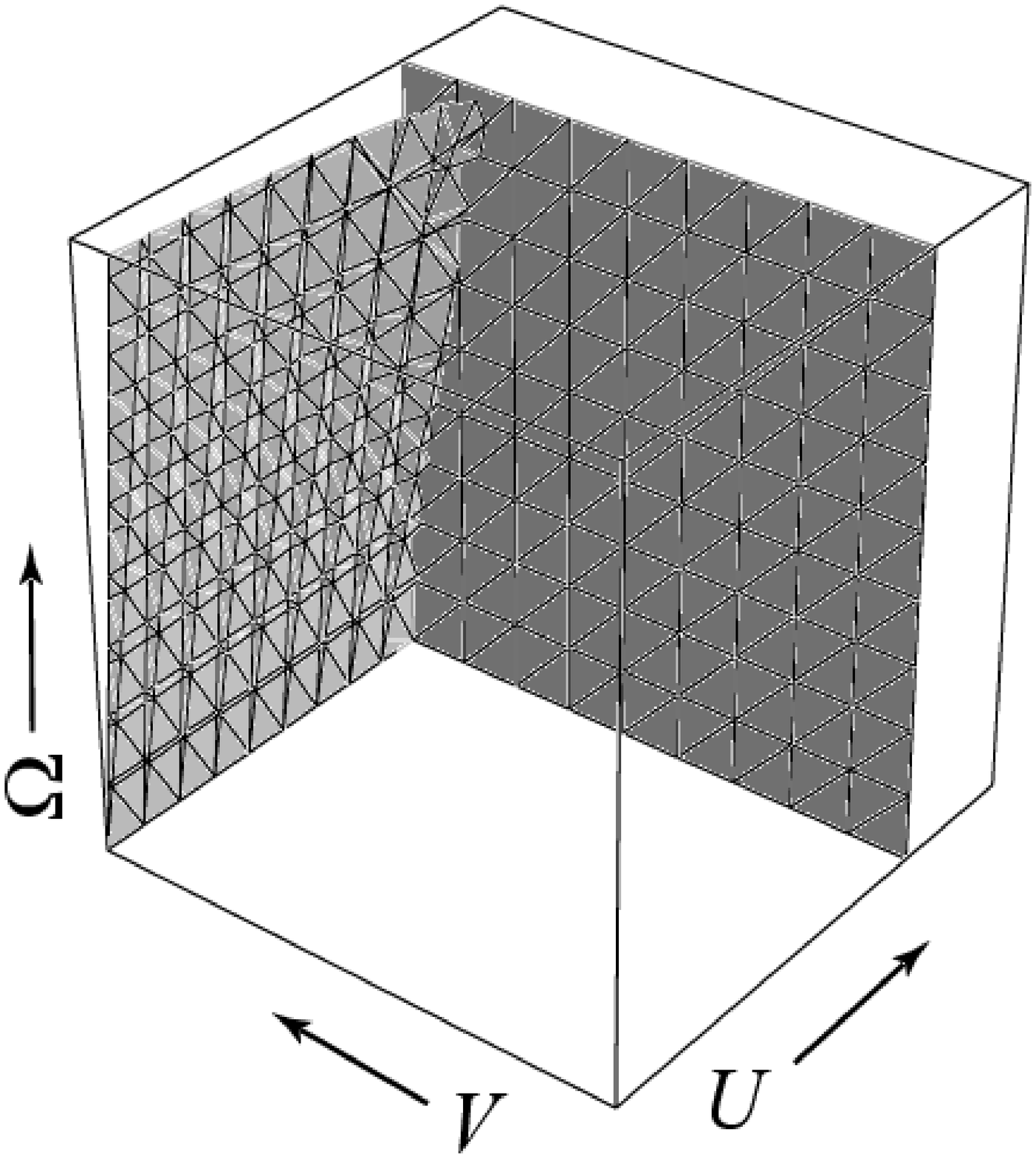}}\quad
        \subfigure[$\Gamma=2$]{
        \label{G20}
        \includegraphics[height=0.35\textwidth]{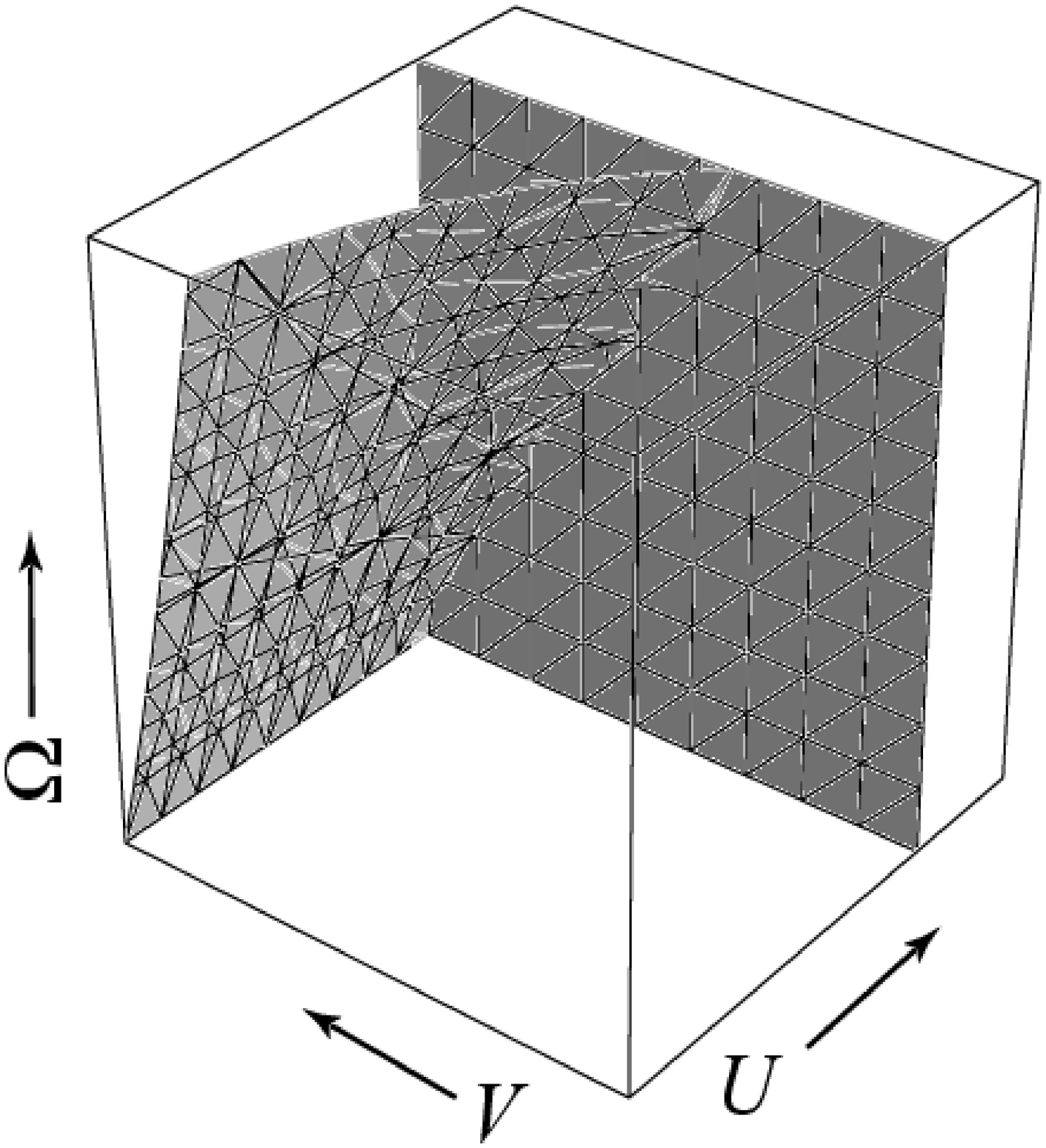}}\quad
    \caption {The Buchdahl surface $\Psi_1=0$ for the relativistic polytrope together with the surface $U=3/4$.}
    \label{fig:Buchsurf}
\end{figure}

By introducing the function
\begin{equation}
\Psi_1:=
1 - \frac{(3{\cal M} + {\cal P})^2}{2(2{\cal M} + {\cal P})}\ ,
\end{equation}
where
\begin{equation}
{\cal M}:=   \frac{Gm}{c^2 r} =
\frac{\sigma(\Omega) V}{1 - V + 2\sigma V}\, ;\quad
{\cal P}:=  \frac{4\pi G}{c^4}r^2 p =
\left(\frac{U}{1-U}\right)
\left(\frac{\sigma^2 V}{1 - V + 2\sigma V}\right)\ ,
\end{equation}
we can write the relativistic Buchdahl inequality as
\begin{equation}
\Psi_1 \geq 0\ .
\end{equation}
Note that the surface inequality ${\cal M} \leq 4/9$ is obtained by
setting ${\cal P}=0$.
The flow on the surface $\Psi_1=0$ is given by
\begin{equation}
\frac{d\Psi_1}{d\lambda}=
\frac{\sigma(3-4U)(1-U+\sigma U)^3 V}{(2(1-U)+\sigma U)^2(3(1-U)+\sigma U)}
\end{equation}
and hence, since $U\leq3/4$ for regular solutions, it follows that
the derivative is positive (zero) when $U<3/4$ ($U=3/4$).
The Buchdahl surface $\Psi_1=0$ and the $U=3/4$ surface are depicted
in Fig.~\ref{fig:Buchsurf} for the
relativistic polytropes with $\Gamma=3/4$ and $\Gamma=2$
(note that the Buchdahl-surface is affected by the equation of state
through $\sigma(\Omega)$).
Since regular solutions start at $V=0$ and
since the flow on the Buchdahl surface is directed toward $V=0$,
the regular solutions cannot pass through the Buchdahl-surface.
Hence, for regular solutions only a limited part of the state
space is accessible (cf.~Fig.~\ref{fig:Buchsurf}).
This constitutes the dynamical systems proof of the Buchdahl inequality.

\section{Translating the state space picture to a physical picture}
\label{sec:trans}

In this section we translate the state space picture to
the common physical variables.
To this end recall, firstly, that an orbit in the interior of the state space
stands for a fluid solution $(m(r), p(r), \rho(r))$ ($r>0$)
inside the fluid body via the transformation~(\ref{uvom}).
Secondly, consider the following {\it auxiliary equations}:
\begin{subequations}\label{UVOmegaaux}
\begin{align}
\label{drdm}
  & \frac{dr}{d\lambda} =  (1-U)(1-V)r\quad,\qquad
  \frac{dm}{d\lambda} = U(1-V)m \\
\label{dpdsig}
&\frac{dp}{d\lambda} = -(1+\sigma)V(1-U+\sigma U)p\quad,\quad
\frac{d\sigma}{d\lambda}=-(1+\sigma)(1-\Upsilon)V(1-U+\sigma U)\sigma \\
\label{mr}
  &\frac{Gm}{c^2 r} =
\frac{\sigma V}{1 - V + 2\sigma V} \\[0.5ex]
\label{implr}
 &r^2 = \frac{c^2}{4\pi G}\left(\frac{U}{1-U}\right)
  \left(\frac{\sigma V}{1 - V + 2\sigma V}\right)
  \rho(\Omega)^{-1} \\
\label{implm}
 &m^2 = \frac{c^6}{4\pi G^3}
  \left(\frac{U}{1-U}\right)
  \left(\frac{\sigma V}{1 - V + 2\sigma V}\right)^3
  \rho(\Omega)^{-1} \:.
\end{align}
\end{subequations}
Based on an equation of state in explicit form~(\ref{rhoexpl}), i.e., $k \rho c^2 = \psi(kp)$,
and a pressure variable $\omega=\omega(kp)$ we can replace $\rho^{-1}$ in the last
two equations by
\begin{equation}
\rho^{-1} = \frac{k c^2}{\psi(\Omega)} \qquad\mbox{or}\qquad
\rho^{-1} = k c^2 \sigma(\Omega) \,\frac{1}{(kp)(\Omega)}\:,
\end{equation}
where $(kp)(\Omega) = \Omega^{1/a}/(1-\Omega)^{1/a}$ in the case $\omega=(kp)^a$.

\begin{figure}[h]
\centering
        \subfigure[$\frac{\ln m}{d\lambda}$]{
        \label{ContourR}
        \includegraphics[height=0.27\textwidth]{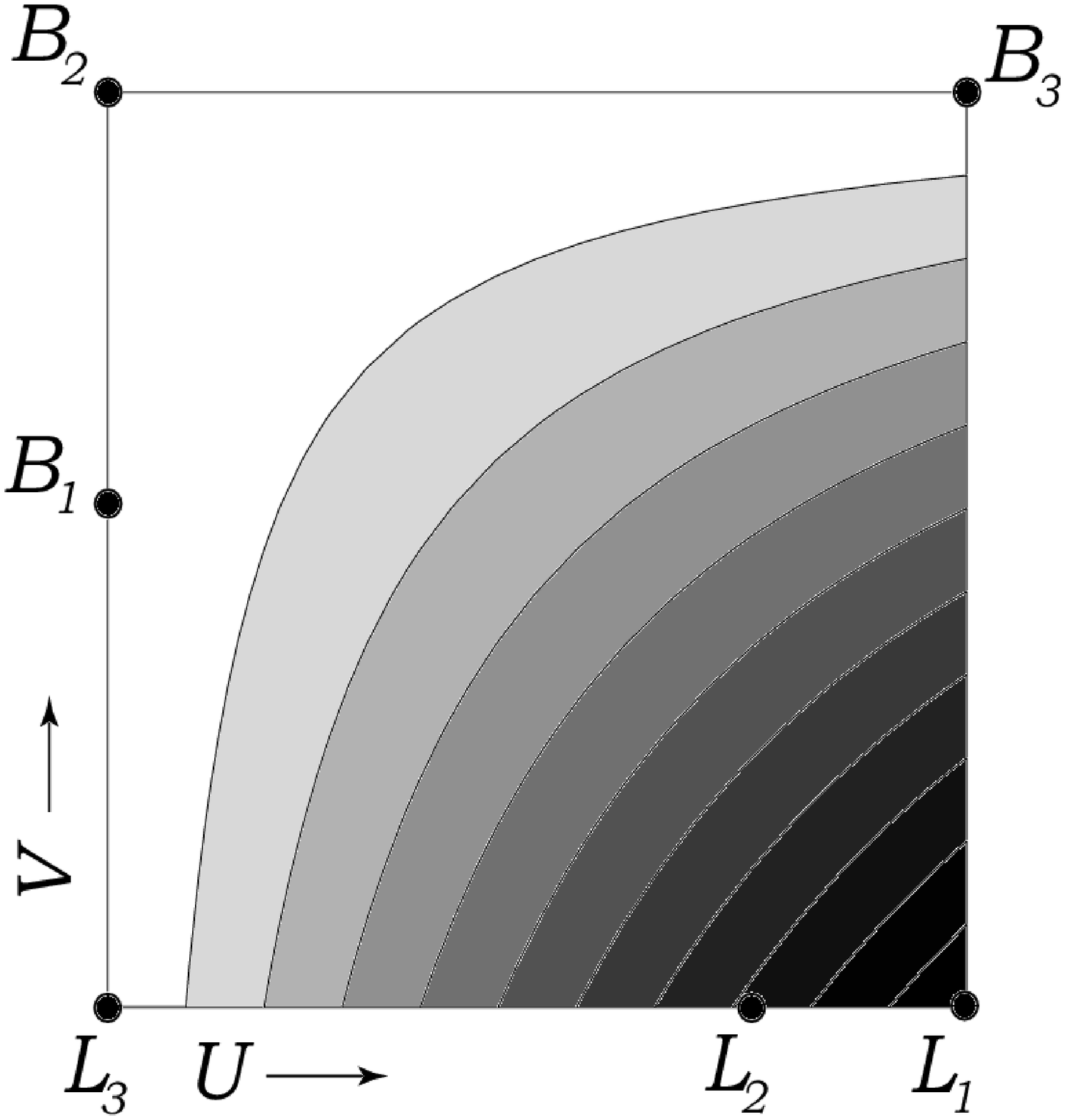}}\quad
        \subfigure[$\frac{\ln r}{d\lambda}$]{
        \label{ContourM}
        \includegraphics[height=0.27\textwidth]{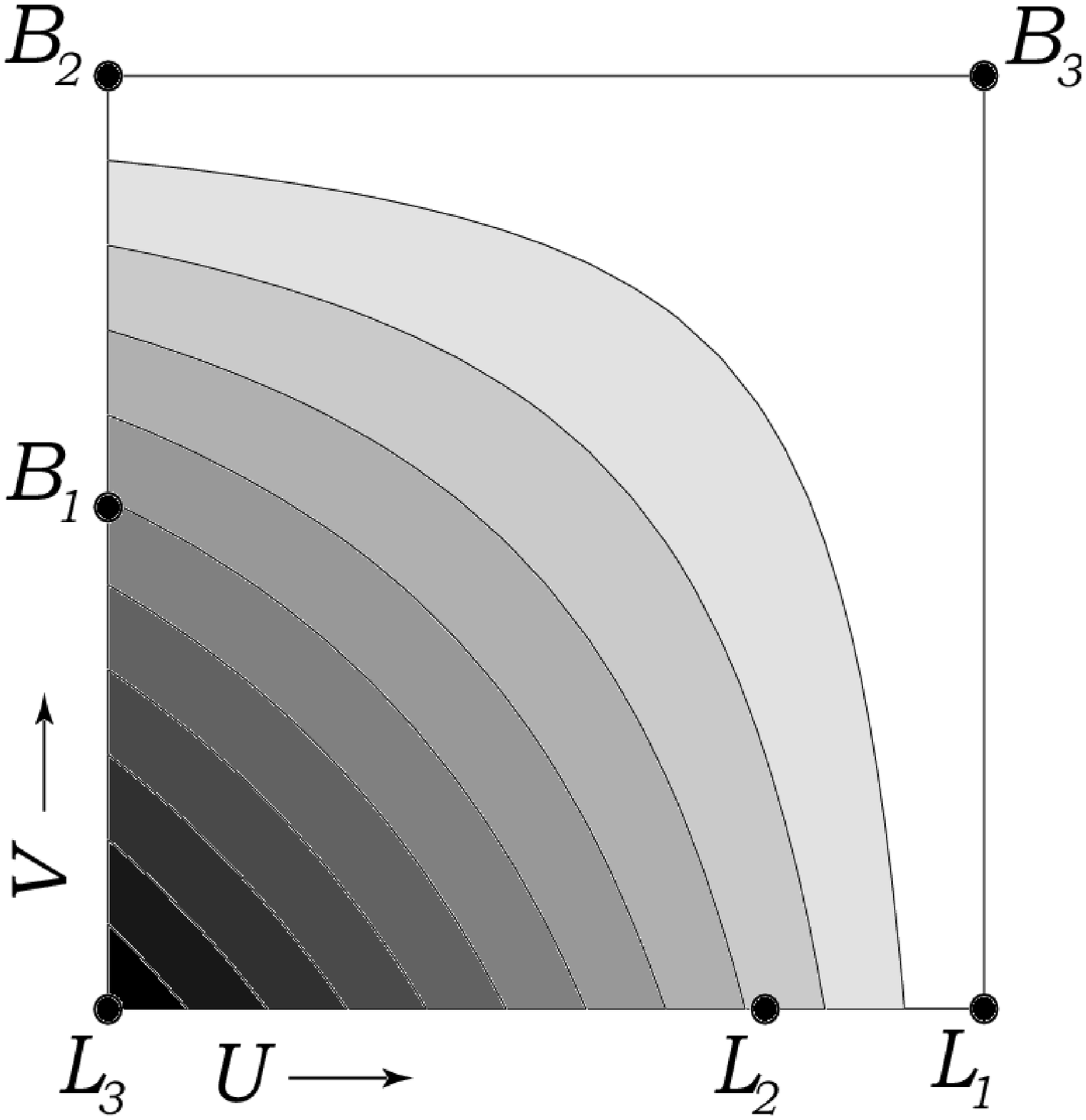}}\quad
    \caption {Contour plots of $\frac{\ln m}{d\lambda}$ and $\frac{\ln r}{d\lambda}$.
      The white regions correspond to
      the minimum value of the functions ($= 0$) and the black to
      the maximum value ($= 1$).}
    \label{fig:Contourplot}
\end{figure}

Equations~(\ref{drdm}) can be used to obtain an intuitive picture of
where solutions gain mass and radius in the state space.
Since these equations are independent of $\Omega$, it is possible
to visualize $d\ln m/d\lambda$ and $d\ln r/d\lambda$ as
contour plots on the $U,V$ plane, see Fig.~\ref{fig:Contourplot}.
The picture suggests that no mass is
acquired near the side faces $U=0$ and $V=1$, and hence the fixed
points $B_1$ and $B_2$ (and $L_5$, if present)
are attractors for solutions with finite
masses. In contrast, $d\ln r/d\lambda$ does not vanish near $B_1$,
but only near $B_2$ (and $L_5$, if present). Therefore, $B_2$ is
the only attractor for solutions with finite radii.
Solutions converging to $B_3$ (or to a periodic orbit $C$ when $n_0=5$)
for $\lambda\rightarrow\infty$
acquire both infinite masses and radii.

To obtain a rigorous derivation of the asymptotic mass and
radius features of the various solutions we must
combine equations~(\ref{UVOmegaaux}) with
the local dynamical systems analysis in the neighborhood of
the fixed points (and the periodic orbits $C$ when $n_0=5$).
We only state the results here (for a detailed discussion of the corresponding
and rather similar Newtonian situation, see \cite{heiugg03}):

Solutions that originate from: (i) $L_1$ are solutions with negative masses
for small $r$; when they have acquired sufficient positive mass,
so that the mass
function $m$ has become zero, they appear in the state space through $L_1$,
cf.~the Newtonian case discussed in \cite{heiugg03});
(ii) $L_2$ have a regular center and are therefore solutions of particular
physical interest;
(iii) $T_3$ have a non-regular center ($\rho \propto r^{-2}$ for $r\rightarrow 0$)
and describe the limit when the central
pressure goes to infinity. One can
show that $T_3$ is associated with a self-similar solution admitting a
$H_4$ spacetime transitive symmetry group.
It is a special case of a solution found by
Tolman~\cite{tol39}, but has been rediscovered many times,
see~\cite{goletal98} and references therein.

Solutions that end at: (i) $B_1$ ($n_0>3$) have infinite radii but
finite masses; (ii) $B_2$ and
(iii) $L_4$ ($n_0=0$) have finite masses and radii and are
therefore the most interesting solutions;
(iv) $B_4$ ($n_0>3$) and (v) $C$ ($n_0=5$) have infinite radii and masses
($m(r) \sim r^{(n_0-3)/(n_0-1)}$ in the case of $B_4$).

Let us now take a closer look at the 2-parameter family of orbits
that converges to $B_2$ as $\lambda\rightarrow\infty$.
All such orbits correspond to perfect fluid solutions with
finite radii $R$ and masses $M$.
To describe the behavior of the physical observables
as $r\rightarrow R$, we will use the local dynamical systems results
together with equations~(\ref{uvom}), (\ref{boundedvar}), or~(\ref{UVOmegaaux}).
Recall that the solutions of the dynamical system~(\ref{UVOmega})
with specified $\Upsilon(\Omega)$ and $\sigma(\Omega)$
represent the totality of perfect fluid solutions
corresponding to a one-parameter family of equations of state, cf.~(\ref{rhoimpl}).
For $p\rightarrow 0$ the leading term of a considered
equation of state was shown to be given by
$k \rho c^2=  s \,(k p)^{\Upsilon_0}$,
where $s$ is a dimensionless constant, cf.~(\ref{eqofstatelow}).

The 2-parameter family of orbits that converges to $B_2$
can be conveniently characterized by the constants
$A$ and $B$, defined according to
\begin{equation}
A:= \lim_{\lambda \rightarrow \infty} U(1-V)^{-n_0}\; ,\quad
B:= \lim_{\lambda \rightarrow \infty}
\Omega^{\frac{1}{f_0(n_0+1)}} (1-V)^{-1}\;.
\end{equation}
Expressed in these quantities we obtain the following, when $r\rightarrow R$,
\begin{subequations}
\label{finitesols}
\begin{align}
p(r) & = k^{-1} B^{n_0+1}\, \delta r^{n_0+1}
+ \frac{n_0+1}{n_0+2}\, k^{-1} B^{n_0+1}
\left(2(n_0+1)+\frac{4\pi R^3}{M c^2}\,k^{-1} B^{n_0+1} A^{-1} (3+2n_0)\right)\delta r^{n_0+2} + O(\cdot) \\
m(r) & = M ( 1- A \delta r^{n_0 +1} ) + O(\cdot)  \\
\rho(r) & = \frac{1}{4\pi R^3}\, M A \,\delta r^{n_0} +O(\cdot)\, ,
\end{align}
\end{subequations}
where $\delta r =\frac{1}{n_0+1} \frac{R-r}{R}$.

The radius $R$ and mass $M$ for a solution are uniquely determined by the values of
$A$ and $B$ through
\begin{equation}\label{AMR}
R^2 = \frac{k c^4}{4\pi s \,G}\:
\frac{A B^{1-n_0}}{s + 2 B}\; ,
\quad\qquad
M^2 = \frac{k c^8}{4\pi s\, G^3}\:
\frac{A B^{3-n_0}}{(s + 2 B)^3}\:.
\end{equation}

The dimensionless quotient $G M/(c^2 R)$ is given by $G M/(c^2 R)
= B/(s+2 B)$, so that the line element~(\ref{ds2}) on the surface
$r=R$ is determined by $e^{2 \Lambda} = c^{-2}e^{-2 \Phi} = 1/(1 -
\sfrac{2G M}{c^2 R}) = 1 + 2 B/s$. (Hence, naturally, $G M/(c^2
R)$ can be used instead of $B$ together with $A$ to parameterize
the different solutions.)

Eq.~(\ref{AMR}) and the related formulas are useful in, e.g.,
numerical computations, since it is fairly easy to compute $A$ and $B$.
Below, in numerical applications, we are going to compare solutions with a
reference solution, i.e., we are going to consider the dimensionless
ratios $R/R_{\rm ref}$, $M/M_{\rm ref}$, where
$R_{\rm ref}$ and $M_{\rm ref}$ are the radius and mass of a typical reference
solution. Of course, the dimensional factors
$(k c^4)/(4\pi s G)$ and $(k c^8)(4\pi s G^3)$ drop out in this case.

For illustrative purposes consider cylindrical coordinates at $B_2$, i.e.,
the coordinate transformation
$(U,V,\Omega) = (\varepsilon \sin\phi, 1-\varepsilon \cos\phi, h^a)$
with $0\leq\phi\leq \pi/2$, $0\leq h$, $0\leq \varepsilon$,
which leads to
\begin{equation}\label{ABcyl}
A = \lim_{\lambda\rightarrow \infty} \epsilon^{1-n_0} \sin\phi\,
(\cos\phi)^{-n_0}\; ,\quad
B = \lim_{\lambda\rightarrow \infty} h^{1/(n_0+1)}\,
(\epsilon \cos\phi)^{-1}\ .
\end{equation}
As every orbit in a neighborhood of $B_2$ converges to $B_2$ as
$\lambda\rightarrow \infty$, we obtain approximate expressions
with arbitrary accuracy
by choosing $\varepsilon$ sufficiently small.
In this picture, every orbit is uniquely characterized by
its intersection point $(h,\phi)$ with the small $\varepsilon$-cylinder,
and accordingly
these values uniquely determine $M$ and $R$ by combining~(\ref{AMR})
with~(\ref{ABcyl}). A similar discussion holds for $L_4$ when $n_0=0$;
for the corresponding Newtonian discussion, see \cite{heiugg03}.

\section{Mass-radius theorems}
\label{mass-radiustheorems}

In this section we will prove several theorems
concerning mass-radius properties of solutions, where
we focus on regular solutions. The underlying equations of state
are as always understood to be
asymptotically polytropic for $p\rightarrow 0$
and asymptotically linear for $p\rightarrow\infty$,
as discussed in Sec.~\ref{equationsofstate}.

\begin{theorem}\label{Upsilonbigger5/6}
All regular solutions have infinite masses and infinite radii
if $\,\Gamma_N \leq \frac{6}{5}$ and $\sigma \leq 1$.
\end{theorem}

\proof
Consider the function $\Psi_2(U,V,\Omega)$, defined by
\begin{equation}\label{psi2}
\Psi_2:= U(8-9V)+7V-6\ .
\end{equation}
The surface $\Psi_2 =0$ coincides with
the regular orbit from $L_2$ to $B_1$ for $n_0=5$ when
projected onto the $\Omega=0$ plane, cf.~(\ref{n5int}).
Taking the derivative of this function and evaluating it
{\it on the surface\/} $\Psi_2 =0$ yields
\begin{equation}\label{dpsi3}
\frac{d\Psi_2}{d\lambda} =
V(1-V)\,\left(\,(5-6\Upsilon)(1-U)^{2} +2\,\sigma\,((4U+U\sigma-1)(3-4U)
-3\Upsilon(1-U)(1+U\sigma))\,\right)\ .
\end{equation}
When $\Gamma_N \leq \frac{6}{5}$ ($\Upsilon \geq \frac{5}{6}$) and $\sigma \leq 1$,
then $d\Psi_2/d\lambda<0$. Since in addition the unstable subspaces
of the fixed points $L_2$ lie in the region $\Psi_2 <0$ of the state space,
it follows that a regular orbit can never leave this region.
The only attractor in this part of the state space is the fixed point $B_4$,
cf.~Theorem~\ref{global}.
Since $B_4$ gives rise to perfect fluid solutions
with infinite masses and radii the claim of the theorem is established.
\proofend

\begin{remark}
The assumption $\sigma \leq 1$ is just the
dominant energy condition, and is satisfied, e.g., for causal relativistic polytropes.
In the theorem, the condition $\sigma\leq 1$ is a sufficient but not necessary;
the statement is valid
for much larger $\sigma$, but not for arbitrarily large values.
\end{remark}

Let us define $\sigma_l:=\left(-4 + \sqrt{16 + 2\,(5\Gamma_N - 6)}\,\right)/6$.

\begin{theorem}\label{Upsilonless5/6}
All regular solutions with
$\Omega_c \in (0,\Omega_{\rm max})$ have finite masses and radii
if $\Gamma_N > \frac{6}{5}$ and if $\sigma < \sigma_l$ on $(0,\Omega_{\rm max})$.
\end{theorem}

\proof
Consider again the function $\Psi_2$ as defined in~(\ref{psi2}).
If $\Gamma_N > 6/5$ (i.e., $\Upsilon < 5/6$),
then $d\Psi_2/d\lambda>0$ is satisfied, provided that
$\sigma < \sigma_l$.
Under the same conditions,
the unstable subspace of a fixed point on $L_2$ lies
in the region $\Psi_2 >0$.
It follows that a regular orbit, which originates from a fixed point $(3/4,0,\Omega_c)$ on $L_2$,
is confined to the region $\Psi_2>0$ of the state space, if
$\sigma(\Omega) < \sigma_l$ for all $\Omega\leq \Omega_c$.
The only attractor in the region $\Psi_2>0$ is the fixed
point $B_2$,
and since $B_2$ generates solutions with finite masses and radii,
the theorem is established.
\proofend

\begin{remark}
In~\cite{Rendall/Schmidt:1991} it is shown that if $6/5<\Gamma_0 <2$
then the solution possesses a finite radius provided that
$\sigma$ is sufficiently small (Theorem 4, p.994).
In~\cite{sim02} and~\cite{art:Heinzle2002}
theorems related to Theorem~\ref{Upsilonless5/6}
are proved with completely different methods.
In the above theorem, the lower limit for $\sigma$ is not
a particularly good lower bound for ``most'' equations of state.
Indeed, as follows from the next theorem, when $n_0 \leq 3$, then
all solutions have finite masses and radii
for all $\Omega_c \in [0,1]$ and thus all values of $\sigma$ are allowed.
\end{remark}

\begin{theorem}\label{n0leq3finite}
(Finiteness of perfect fluid solutions).
All regular and non-regular perfect fluid solutions
have finite radii and masses if $n_0 \leq 3$ (i.e., $\Gamma_0 \geq 3/4$).
\end{theorem}

\proof
This theorem follows from Theorem~\ref{n03}, which shows that
all solutions end at $B_2$ if $n_0\leq3$ (at $L_4$ if $n_0=0$),
and from that $B_2$ ($L_4$) is associated with
solutions with finite masses and radii.
\proofend

\begin{remark}({\it General validity}).
The last theorem holds irrespective of the asymptotic behavior
of the equation of state at high pressures.
Also for the first two theorems the
asymptotic high pressure regime is irrelevant
if one restricts the attention to solutions with finite central pressures.
This is because the proofs in such cases do not rely on the inclusion
of the boundary $\Omega=1$ in the state space.
This is in contrast to the next theorem which makes use
of the $\Omega=1$ subset and relies on our assumptions about
differentiability of the equation of state when $\Omega\rightarrow 1$ ($p \rightarrow \infty$).
\end{remark}

\begin{theorem}\label{spiralthm}(Spiral structure of the $(M,R)$-diagram).
Let $0<n_0\leq 3$.%
\footnote{More generally, we can consider
  equations of state with arbitrary $n_0$
  as long as the orbit that originates from $T_3$ ends at $B_2$
  (or $L_4$ if $n_0=0$).
  }
For sufficiently high central pressures
the mass-radius diagram exhibits a spiral structure,
i.e., $(R(p_c), M(p_c))$ is given by
\begin{equation}\label{spirals2}
\left(\begin{array}{c}
R(p_c) \\
M(p_c)
\end{array}\right) \:=\:
\left(\begin{array}{c}
R_O \\
M_O
\end{array}\right)
\:+\:
\left(\frac{1}{p_c}\right)^{\tau_1} \: {\mathcal B} \:
{\mathcal J}(\tau_2 \log p_c\,)\: b
\:+\:
o(\left(\frac{1}{p_c}\right)^{\tau_1}) \:,
\end{equation}
where $R_O$ and $M_O$ are constants, ${\mathcal B}$ is a
matrix with positive determinant, and $b$ a non-zero vector.
The matrix ${\mathcal J}(\varphi) \in \mathrm{SO}(2)$ describes a
(positive) rotation
by an angle $\varphi$, and the constants $\tau_1$ and $\tau_2$ are given by
\begin{equation}\label{gamma122}
\tau_1 = \frac{3 \gamma_1-2}{4 \gamma_1}\, , \qquad
\tau_2 = \frac{1}{4 \gamma_1}\, \sqrt{c}\:,
\end{equation}
where $\sqrt{c} = \sqrt{-36 +44 \gamma_1 - \gamma_1^{\:2}}$
see Table~\ref{tab:UVcube}.
\end{theorem}

{\it Sketch of proof}.
The basic observation is that the regular orbit on $\Omega=1$
forms a spiral as it converges to $T_3$, and that this spiral
subsequently reflects itself as a spiral in the $(R,M)$-diagram.

Choose $\varepsilon$ small and set $\omega_\varepsilon = 1-\varepsilon$.
The dynamical system asymptotically decouples as $\Omega\rightarrow 1$
(recall from~(\ref{Upsilonachievement}) and~(\ref{sigmaachievement})
that we can always choose $\omega$ in such a way that
$\frac{d\Upsilon}{d\Omega} = O((1-\Omega)^k)$
and $\frac{d\sigma}{d\Omega} = O((1-\Omega)^k)$ as $\Omega\rightarrow 1$),
which enables us to approximately solve the system
on $[0,1]^2 \times [\Omega_\varepsilon,\Omega]$:
let $U_{\mathrm{L}}(\lambda)$, $V_{\mathrm{L}}(\lambda)$
denote the regular solution of the
system~(\ref{Ueq}) and (\ref{Veq}) with $\Gamma_N=1$, $\sigma=\sigma_1=\gamma_1-1$.
Then $(U_{\mathrm{L}}(\lambda), V_{\mathrm{L}}(\lambda), \Omega(\lambda))$
is an approximate solution of~(\ref{UVOmega}) when
$\frac{d\Omega}{d\lambda} = -f_1 (1-\Omega) H_{\mathrm{L}}$,
where $H_{\mathrm{L}} = H(U_{\mathrm{L}},V_{\mathrm{L}},1)$.

For large $\lambda$ the orbit $(U_{\mathrm{L}}(\lambda), V_{\mathrm{L}}(\lambda))$
has the form of a spiral, i.e.,
\begin{equation}\label{hatspiral}
\left(\begin{array}{c}
U_{\mathrm{L}}(\lambda)\\
V_{\mathrm{L}}(\lambda)
\end{array}\right) \:=\:
\left(\begin{array}{c}
U_{T_3}\\
V_{T_3}
\end{array}\right)
+ \exp(-\delta_1 \lambda) \: {\mathcal B}^\prime \:
{\mathcal J}(\delta_2 \lambda)\: b^\prime\:,
\end{equation}
where $U_{T_3}, V_{T_3}$ are the coordinates of the
fixed point $T_3$; ${\mathcal B}^\prime$ is a
matrix with $\det {\mathcal B}^\prime >0$, and $b^\prime$ a non-zero vector;
$\delta_1 = \frac{\gamma_1}{4(\gamma_1^2+2)} (3\gamma_1-2)$
and $\delta_2 = \frac{\gamma_1}{4(\gamma_1^2+2)}\, \sqrt{c}$.
The constants $\delta_1$ and $\delta_2$
are the real and imaginary part of the complex eigenvalue
that is associated with $T_3$,
cf. with Table~\ref{tab:UVcube}.

Inserting~(\ref{hatspiral}) into the equation for $\Omega$ we obtain the
approximation
\begin{equation}\label{omegaspiralsim}
1-\Omega(\lambda) =
(1-\Omega_c)\, const\, \exp\big(f_1 H_{T_3} \,(\lambda-\tilde{\lambda})\big),
\end{equation}
where $H_{T_3} = H(U_{T_3}, V_{T_3},1) = \gamma_1^2/(2+\gamma_1^2)$,
and $\tilde{\lambda}$ and $const$ are independent of $\Omega_c$
(the value of $\Omega(\lambda)$ as $\lambda\rightarrow -\infty$).

Consider the orbit that originates from $T_3$ into the state space:
it intersects the plane $\Omega=\Omega_\varepsilon$ in a point
$I$ with approximately the same $(U,V)$-coordinates as $T_3$.
A regular solution with $\Omega_c$ sufficiently close to $1$
can be described by~(\ref{hatspiral}) and~(\ref{omegaspiralsim}).
It intersects $\Omega=\Omega_\varepsilon$ at $\lambda=\lambda_c$,
where
\begin{equation}\label{tc}
\lambda_c = \frac{1}{a H_{T_3}}\,\log\frac{1}{1-\Omega_c}+const =
\frac{1}{H_{T_3}}\log p_c + const\:.
\end{equation}
For the components $(U(\lambda_c), V(\lambda_c))$ of this regular solution
we have $(U(\lambda_c), V(\lambda_c)) = (U_{T_3},V_{T_3}) + (\delta U_0,\delta V_0)$
with
\begin{equation}\label{censpiral}
\left(\begin{array}{c}
\delta U_0(\eta_c)\\
\delta V_0(\eta_c)
\end{array}\right) \:=\:
\left(\frac{1}{p_c}\right)^{\tau_1}  \: {\mathcal B}^{\prime\prime} \:
{\mathcal J}(\tau_2 \log p_c\,)\: b^{\prime\prime}\:,
\end{equation}
where ${\mathcal B}^{\prime\prime}$ ($\det{\mathcal B}^{\prime\prime} >0$)
and $b^{\prime\prime}$ have absorbed the constants.
From~(\ref{censpiral}) we see that the set of regular solutions
intersects $\Omega=\Omega_\varepsilon$ in a spiral
in a neighborhood $U_I$ of $I$ in this plane.

The orbit that originates
from $T_3$ and passes through $I\in U_I$
eventually intersects the mass-radius cylinder (cf.~(\ref{ABcyl}))
at some point $O\in U_O\subseteq \mathrm{cylinder}$,
where the cylindrical coordinates of $O$ determine the
mass $M_O$ and the radius $R_O$ of the solution (cf.~(\ref{AMR})).
The flow of the dynamical system induces a diffeomorphism
$U_I \rightarrow U_O$, so that the spiral~(\ref{censpiral}) in $U_I$ is
mapped to a distorted (positively oriented) spiral in $U_O$.
Since the cylindrical coordinates $(\phi,h)$ in the
small neighborhood $U_O$
are related to $(R,M)$ by a positively oriented linear
map, the spiral~(\ref{censpiral}), with different ${\mathcal B}$ and $b$,
appears also in the $(R,M)$ diagram.
\proofend

\begin{remark}
Note that the theorem
does not rely on the behavior of the equation of state in the intermediate
non-asymptotic regimes. Hence the theorem describes a universal phenomenon
and connects it with the self-similar solution that corresponds to $T_3$.
A similar theorem has been proved in~\cite{art:Makino2000} using quite
different methods. However, an advantage with the present approach, apart
from less restrictive assumptions, brevity of proof, and clarity, is that
it visually shows the importance of key solutions like the self-similar
one associated with $T_3$.
\end{remark}

\section{Examples}
\label{sec:ex}

In this section the dynamical systems approach to relativistic stellar models
will be illustrated by several examples. The presented results also demonstrate
the usefulness of the new framework in numerical computations.

\subsection{Relativistic polytropes}
\label{subsec:relpol}

\begin{figure}[h]
\centering
        \subfigure[$\Gamma=4/3$]{
        \label{RM1}
        \includegraphics[height=0.27\textwidth]{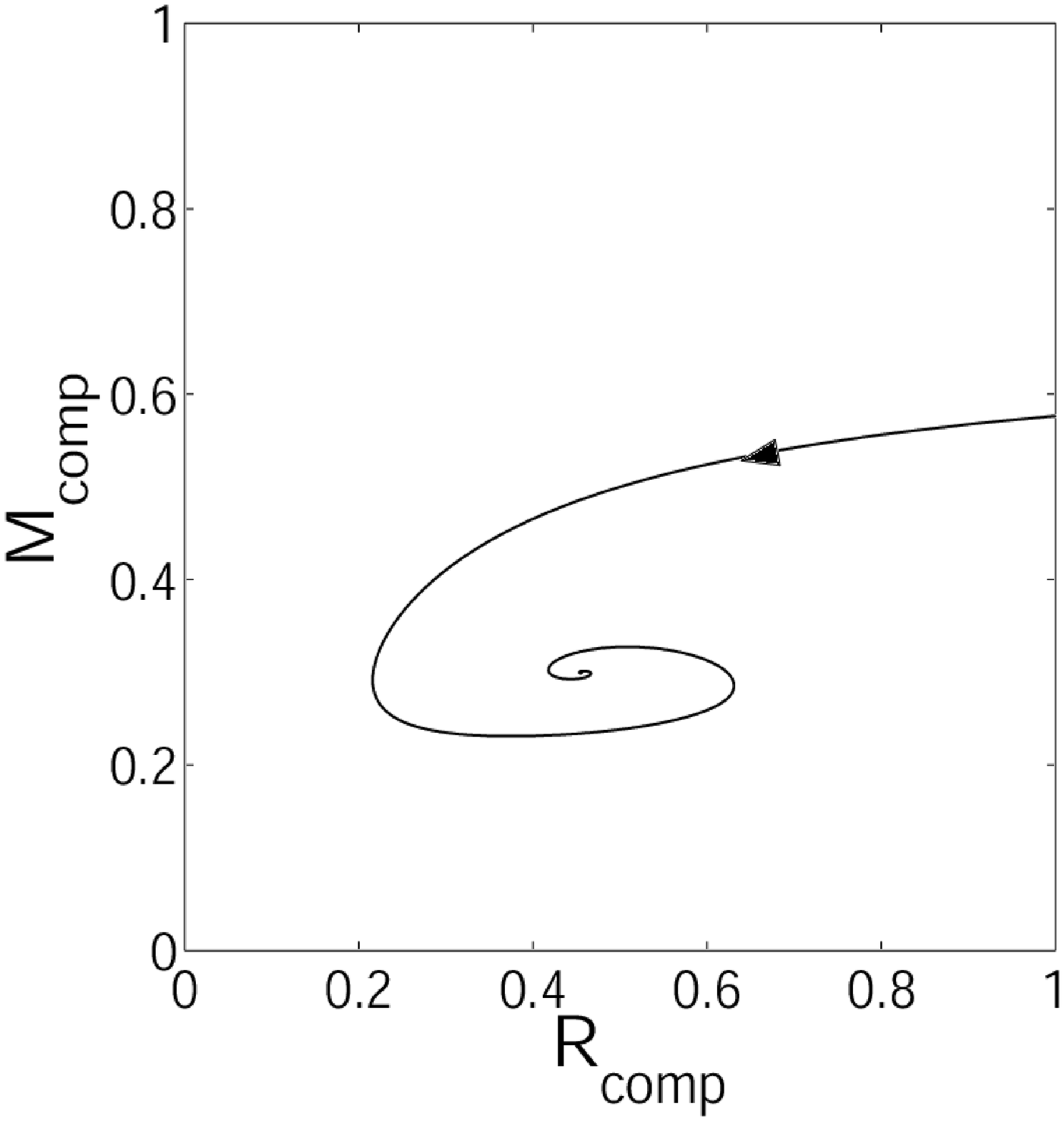}}\quad
        \subfigure[$\Gamma=5/3$]{
        \label{RM2}
        \includegraphics[height=0.27\textwidth]{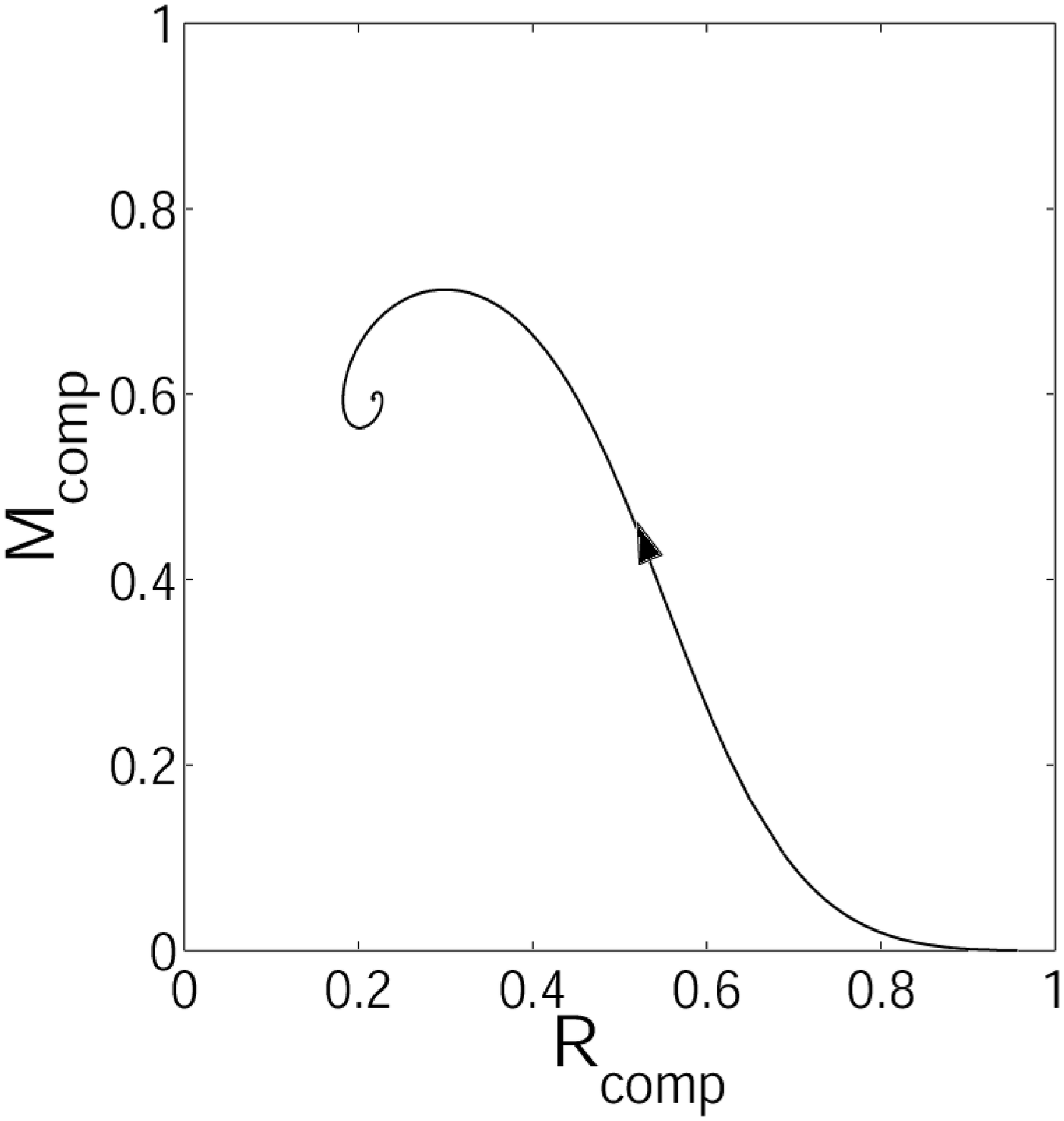}}\quad
        \subfigure[$\Gamma=2.0$]{
        \label{RM3}
        \includegraphics[height=0.27\textwidth]{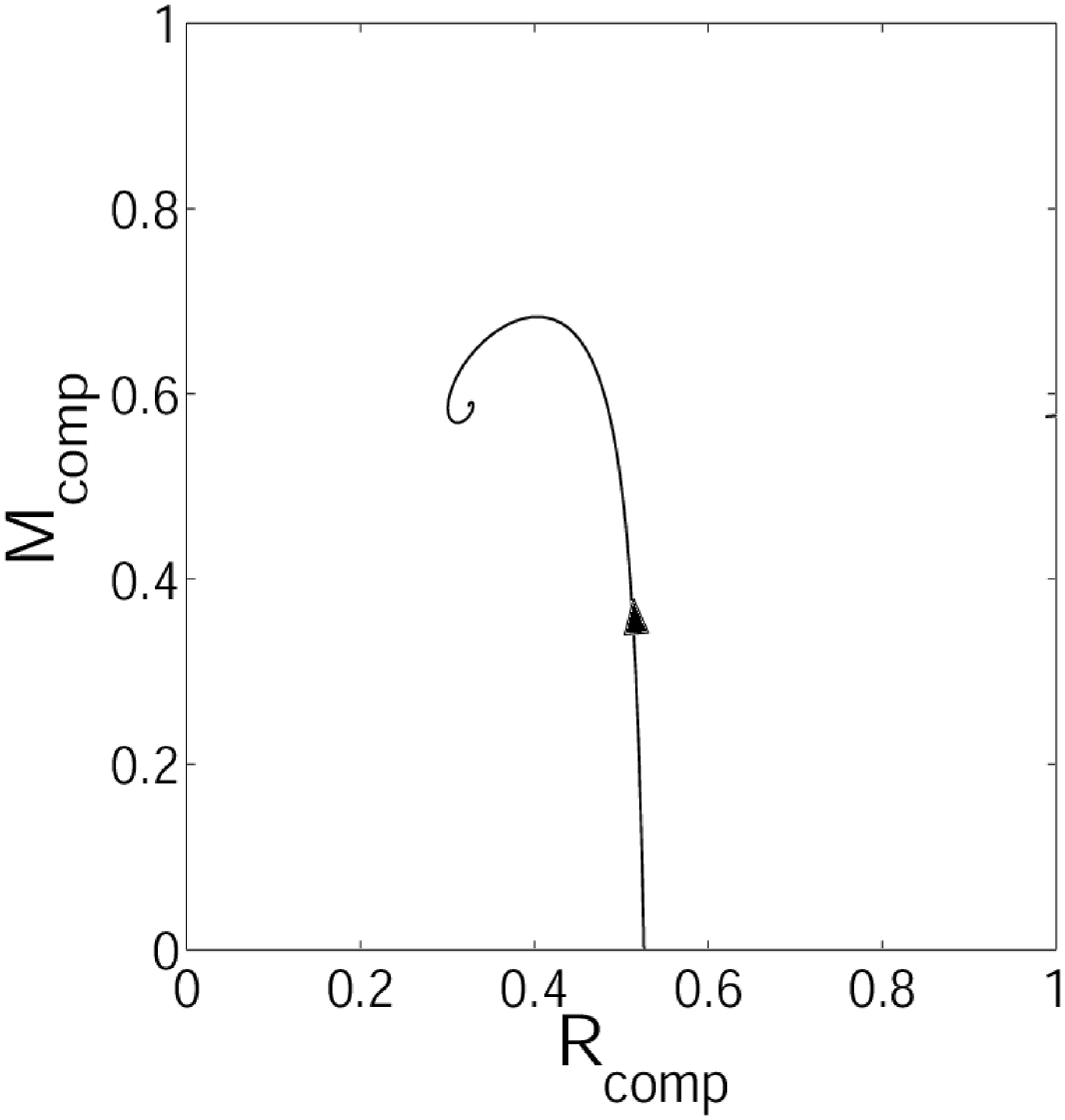}}\quad
    \caption{Mass-radius diagrams for the regular relativistic polytropes. We display the
      compactified quantities
      $M_{{\rm comp}}=(M/M_{{\rm ref}})/(1+M/M_{{\rm ref}})$ and
      $R_{{\rm comp}}=(R/R_{{\rm ref}})/(1+R/R_{{\rm ref}})$,
      where $M_{{\rm ref}}$ and $R_{{\rm ref}}$ are typical values.
      The arrows indicate the direction of increasing $\Omega_c$, i.e.,
      increasing central pressures.}
    \label{fig:MRdiagramrelpol}
\end{figure}

As discussed in Sec.~\ref{equationsofstate}, a relativistic polytrope
with index $\Gamma$ can be described by the dynamical system~(\ref{UVOmega})
with $\sigma = (\Gamma-1) \Omega$ and $\Upsilon =  (1 + (\Gamma - 1)\Omega)/\Gamma$.
The numerical integration of this system is straight forward,
when we use the local analysis near the fixed points.
The radius $R$ and the total mass $M$ of a
solution can be obtained by various methods,
in particular one can use equations~(\ref{implr}) and~(\ref{implm}), or
one can include the auxiliary equations~(\ref{drdm}) in the numerical
integration of
the dynamical system; Eq.~(\ref{AMR}) provides a different approach.
Mass-radius diagrams for the regular solutions associated with different values
of $\Gamma$ are shown in Fig.~\ref{fig:MRdiagramrelpol}.

The theorems of Sec.~\ref{mass-radiustheorems} guarantee that
all regular solutions possess finite $R$ and $M$ for $n_0\leq 3$ ($\Gamma\geq 4/3$).
For $5> n_0 > 3$ ($\Gamma<4/3$), regular solutions with small central
pressures must be finite fluid bodies.
The numerical investigation improves these results: all regular solutions
associated with $\Gamma \gtrsim 1.2971$ ($n_0 \lesssim 3.366$)
possess finite $R$ and $M$.
A similar analysis was performed in~\cite{nilugg00} for purely
polytropic equations of state:
for a polytropic index $n \lesssim 3.339$ only finite fluid bodies appear.
For an extensive discussion of the methods involved in establishing
these results we refer the reader to that paper.

\begin{figure}[h]
\centering
        \subfigure[$\Gamma=1.1$]{
        \label{G11}
        \includegraphics[height=0.26\textwidth]{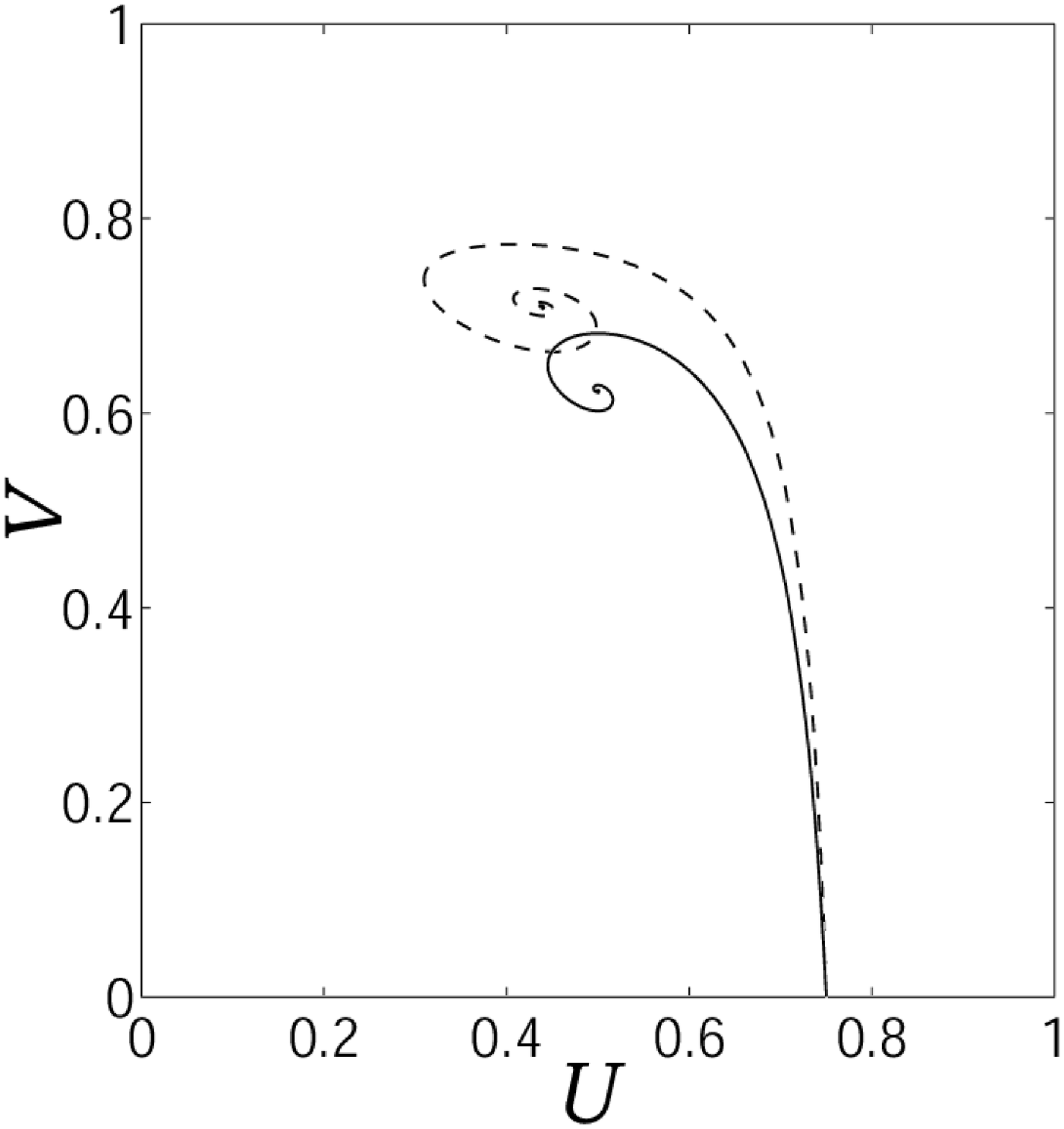}}\quad
        \subfigure[$\Gamma=1.01$]{
        \label{G101}
        \includegraphics[height=0.26\textwidth]{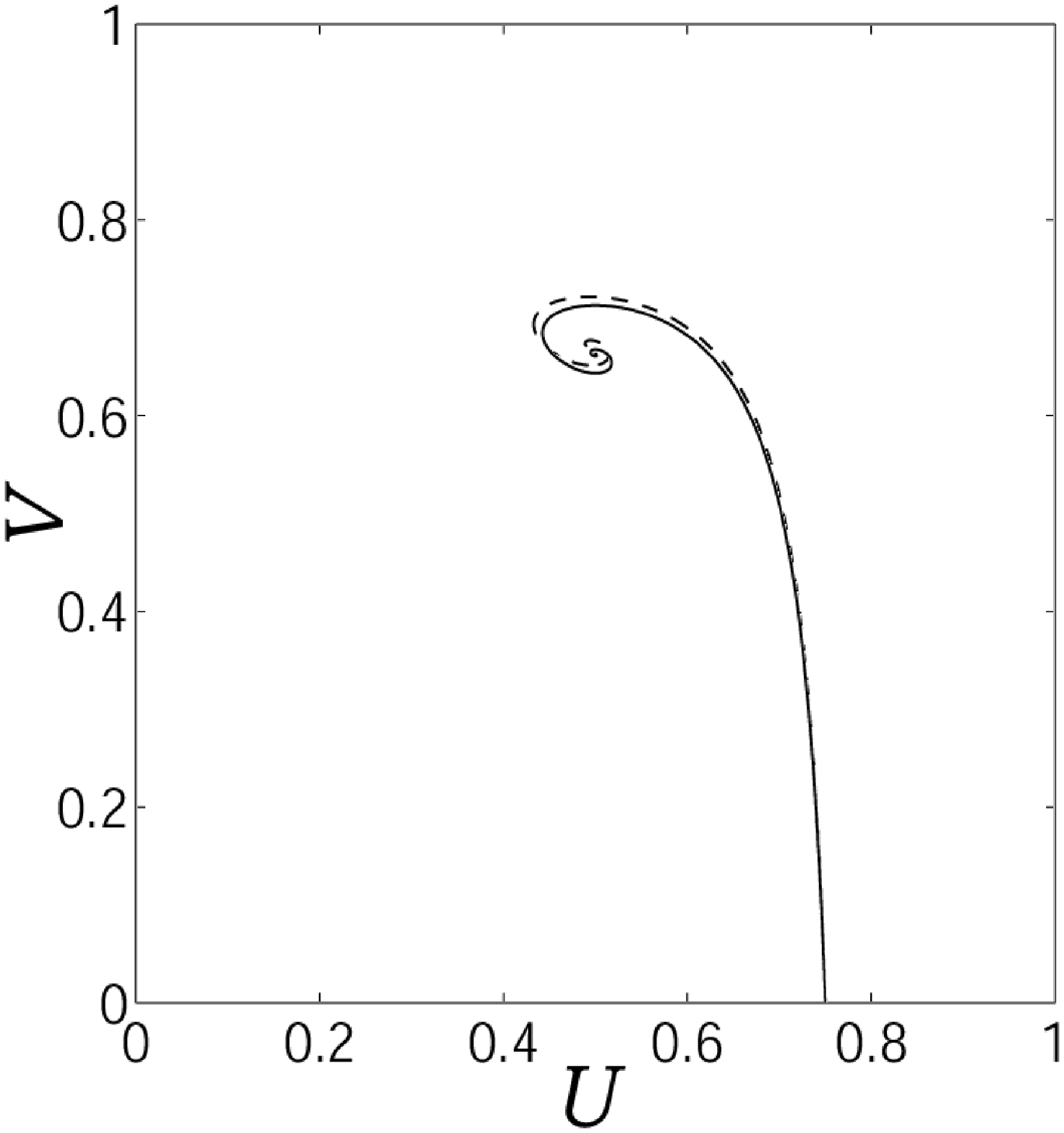}}\quad
        \subfigure[$\Gamma=1.001$]{
        \label{G1001}
        \includegraphics[height=0.26\textwidth]{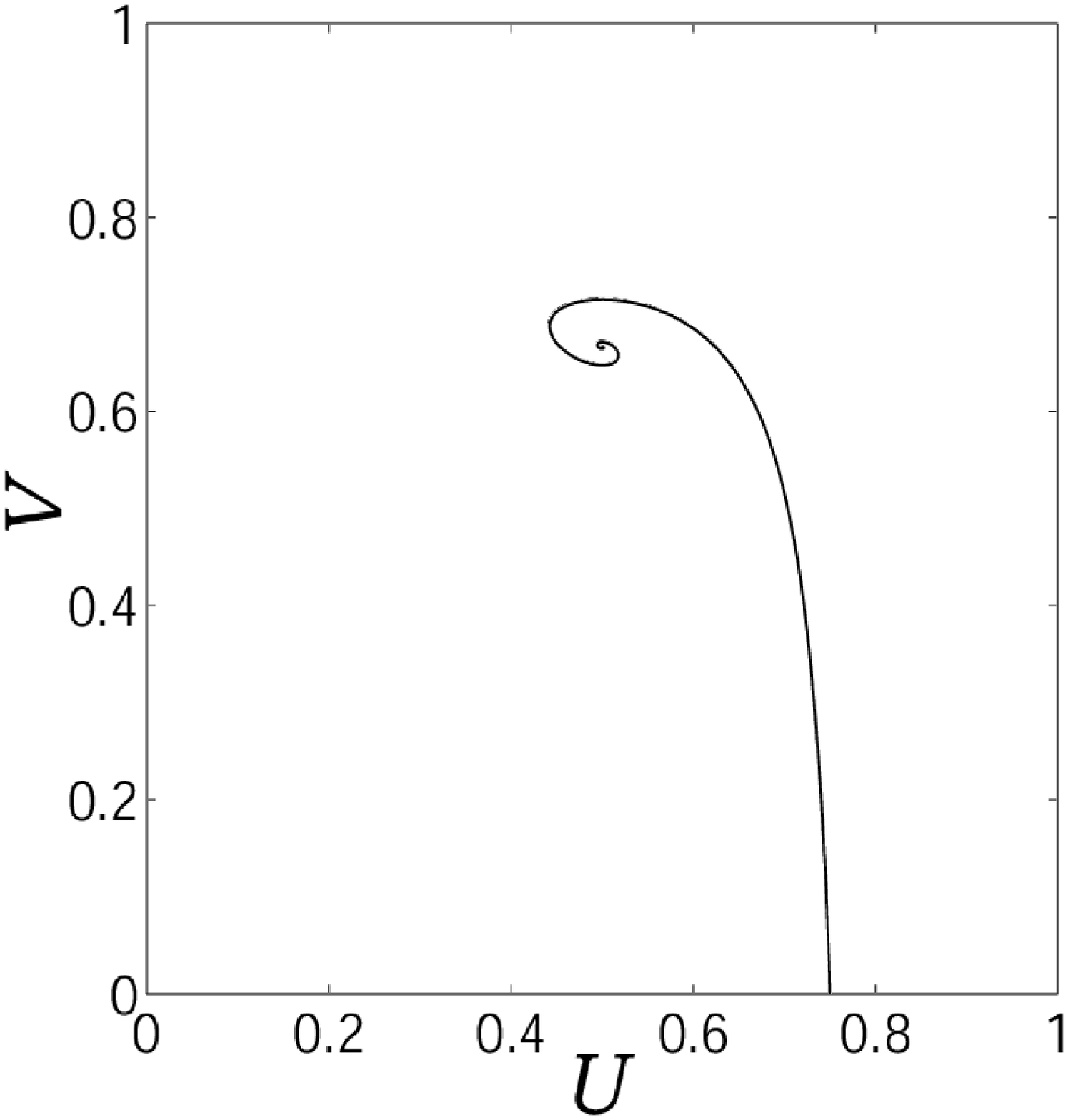}}\quad
    \caption{Superimposed regular orbits for the relativistic polytrope,
      with varying $\Gamma=\Gamma_0=\gamma_1$.
      The dashed line corresponds to the orbit in the $\Omega = 0$
      plane and the solid line to the orbit in the $\Omega=1$ plane.}
    \label{fig:superimp}
\end{figure}

Letting $\Gamma \rightarrow 1$ for relativistic polytropes implies that
$\sigma \rightarrow 0$, and hence the dynamical system~(\ref{UVOmega})
approaches the Newtonian system, cf.~Appendix~\ref{A}.
Moreover, the equation for $\Omega$ decouples from
the $U$- and $V$-equations, and the flows on the $\Omega=0$ and
the $\Omega=1$ subset coincide in the limit.
It follows that the regular surface is a surface perpendicular
to the $\Omega=const$ plane; its projection to $\Omega=const$ is
depicted in Fig.~\ref{G1001}.
Note also that the limit $\Gamma\rightarrow 1$ reveals a
relationship between the Newtonian self-similar
solution associated with $B_4$ (discussed, e.g., in~\cite{Chandrasekhar:1939})
and the relativistic self-similar solution associated with $T_3$.

Regular solutions associated with an equation of state characterized by
an arbitrary adiabatic index $\Gamma(p)$ close to $1$ lie
close to the regular "soft-limit-surface".%
\footnote{Consider an arbitrary one-parameter class of equations of state
  characterized by a parameter $\mu>0$, where
  $\Gamma_0$ and $\gamma_1$ converge to $1$ as $\mu\rightarrow 0$.
  Then the associated regular surface approaches the soft-limit-surface
  as $\mu\rightarrow 0$.}
Hence, deviations from the soft-limit-surface
measure both deviations of the equation of state from
the soft limit and relativistic effects.
Stiffer models are located "further out" compared to the
soft-limit-surface and increasing stiffness implies that
the $U=3/4$ surface is approached, and it is on this
``stiff-limit-surface'' the incompressible perfect fluid
solutions reside, as discussed previously.
Note also that the Buchdahl surface, depicted in Fig.~\ref{fig:Buchsurf},
for increasingly soft equations of state moves toward the $V=1$ surface,
thereby becoming increasingly less restrictive. This
illustrates that the Buchdahl inequality is a purely relativistic effect,
since the surface disappears from the interior state space in the
soft, and thereby Newtonian, limit.


\subsection{The ideal neutron gas}
\label{subsec:neutron}

The equation of state of the ideal neutron gas provides a simple
model for neutron star matter; it is given in implicit form
in equation~(\ref{neutron}).
As discussed in Sec.~\ref{equationsofstate}, the ideal neutron gas
can be naturally described in the dynamical systems framework,
and is easy to handle numerically.

\begin{figure}[h]
\centering
\includegraphics[height=0.35\textwidth]{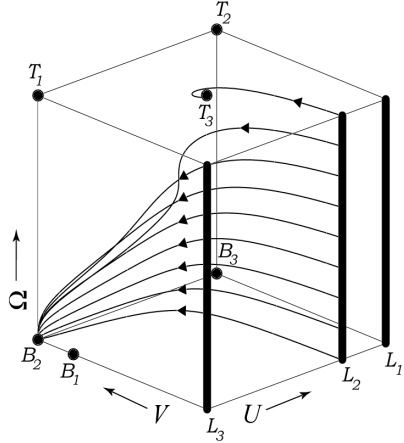}
\caption {Typical regular orbits for the ideal neutron gas
together with the orbits on the boundaries $\Omega=0$ and $\Omega=1$.}
\label{neutronorbits}
\end{figure}

In Fig.~\ref{neutronorbits} regular orbits
for various values of $\Omega_c$ and the regular orbits
on $\Omega=0$ and $\Omega=1$ are depicted.

\subsection{Composite equations of state}
\label{subsec:composite}

Consider first a stiff perfect fluid characterized by the equation
of state $p=(\rho-\rho_-)c^2$, which can be rewritten as
\begin{equation}
k\rho c^2 = 1 + k p\:,
\end{equation}
where $k=(\rho_- c^2)^{-1}$.
If we choose $\omega = k p$, we
obtain $f=1$, $\Upsilon=\Omega$, and $\sigma =\Omega$,
i.e., the dynamical system~(\ref{UVOmega}) becomes very simple.
In this context, it is of interest
to note that the solution that corresponds to the
orbit from $T_3$ to $L_4$ (which is a solution with finite radius and mass
but infinite central pressure) is explicitly known: it
is a special case of a Tolman solution~\cite{tol39}.
In the present formalism,
the corresponding orbit is given by $U=3/(5+\Omega)$ and $V=(2+\Omega)/(2+3\Omega+4\Omega^2)$.
(For interesting features
of this solution, see \cite{karetal02}.)

As an example of a composite equation of state,
let us continuously join a stiff equation of state
$p = (\rho-\rho_-)\, c^2$ (when $p>p_j$) to a
relativistic polytrope (when $p<p_j$).%
\footnote{If an equation of state is only known for low pressure
  (as is true in reality), then mass estimates for stellar models
  can be obtained by extending the equation of state as
  a stiff fluid, see e.g.,~\cite{har78}.}
Such an equation of state possesses a kink at $p_j$.

\begin{figure}[h]
\centering
\includegraphics[height=0.40\textwidth]{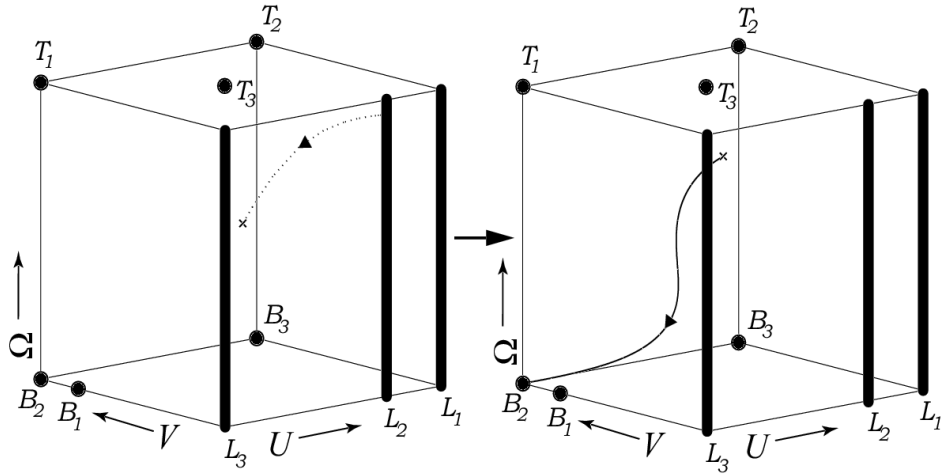}
\caption {An example of an orbit for a
  composite equation of state consisting of a relativistic polytrope
  ($\Gamma=11/6$) and a stiff fluid.
  We have chosen to cut the "stiff fluid orbit" at $\Omega=0.50$ (where $(U,V)=(0.58,0.60)$).
  The jump transformation law then yields a map to the
  point $(U,V,\Omega)=(0.58,0.60,0.75)$ in the state space of the
  relativistic polytrope, which serves as initial data for the subsequent evolution in that
  state space.}
\label{composite}
\end{figure}

As outlined in Sec.~\ref{equationsofstate}, we work
with two state spaces and dynamical systems, one
pertaining to the relativistic polytrope and one
representing the stiff fluid.
The jump at $p_j$ reflects itself in a map between the two state spaces.
Since $\rho$ is continuous, the values of $U,V$ are continuous under
the map, however, there is a jump in $\Omega$.
Since $\rho_{{\rm relpol}}(p_j) = \rho_{{\rm stiff}}(p_j)$ we have
\begin{equation}
\frac{1}{\Gamma-1} \,\left(\frac{(kp_j)^{1/\Gamma}}{k} + p_j\right) =
\rho_- c^2 + p_j\:.
\end{equation}
Dividing by $p_j$ and using that $\omega_{{\rm relpol}}(p) = (k p)^{(\Gamma-1)/\Gamma}$
and $\omega_{{\rm stiff}}(p) = (\rho_- c^2)^{-1} p$, we
obtain that
\begin{equation}
\Omega_{{\rm relpol}}=\frac{\Omega_{{\rm stiff}}}{\Gamma-1}
\end{equation}
at the jump.
If $\Gamma=2$, i.e., if the relativistic polytrope is asymptotically stiff,
then there is no jump at all.
A typical solution with a jump is illustrated in Fig.~\ref{composite}.

Note that, in general, if one starts
with a regular solution in the first state space,
then this solution has to be matched with a
non-regular one associated with the other (extended)
equation of state in the second state space.

\section{Concluding remarks}
\label{sec:concl}

In this paper we have derived a dynamical systems formulation
for the study of spherically symmetric relativistic stellar models.
The method of "homology invariants"
known from the theory of Newtonian polytropes has been generalized both to
a general relativistic context and to the broad class
of asymptotically polytropic/linear equations of state.
For previous related work see~\cite{heiugg03} (Newtonian asymptotic polytropes)
and~\cite{nilugg00} (polytropes in GR).

The present formulation has turned out to be
advantageous in many respects:
(i) it provides a clear visual representation
of the solution spaces associated with broad classes of
equations of state; in particular it is revealed how
the global qualitative properties of the solution
space are influenced by the equation of state;
(ii) the formulation makes the theory of dynamical systems available,
which makes it possible to prove a number of theorems
and describe the qualitative behavior of solutions;
and (iii) the framework is particularly suited for
numerical computations, since the numerics is supported by
the local dynamical systems analysis.

The idea of using dimensionless variables and exploiting
asymptotic symmetries and properties to derive a dynamical systems
formulation is a quite versatile one. As another example, in a
future paper, we will show how one also can treat a collisionless
gas. The "dynamical systems approach" can even be applied to
problems, in general relativity and other areas, when no
symmetries exist at all \cite{uggetal03}, as will be shown in
another set of papers. Thus the ideas in the present work should
be seen in a quite broad context, and it is likely that there are
many problems in very different areas that could benefit from the
type of ideas and the approach presented in this paper.

\subsection*{Acknowledgements}
The authors would like to thank Alan Rendall and Robert Beig
(J.M.H.) for helpful discussions. This work was supported by the
DOC-program of the {\it Austrian Academy of Sciences} (J.M.H.) and
the {\it Swedish Research Council} (C.U.).

\begin{appendix}
\section{Appendix: Comparison with the Newtonian theory}
\label{A}

The Newtonian equations of hydrostatic equilibrium are obtained
from~(\ref{potential}) and (\ref{oppvol}) by letting $c\rightarrow \infty$, however,
note that in the Newtonian case $\rho$ is interpreted as the rest-mass density
and $c^2 \Phi \rightarrow \Phi_N$, the Newtonian potential.
Consistently, since $\sigma = p/(\rho c^2)$,
in the dynamical system~(\ref{UVOmega})
we must set $\sigma \equiv 0$. Note also that $v$ reduces to $v= G\rho m/(r p)$.
Since $\sigma\rightarrow 0$ for $p\rightarrow 0$, the relativistic
system~(\ref{UVOmega}) and its Newtonian counterpart coincide on $\Omega=0$.

In the Newtonian case we can treat equations of state $\rho(p)$ with less restrictive
asymptotic behavior in the regime $p \rightarrow \infty$. This is because
it was the existence of $\sigma(\Omega)$ on the right hand side of the dynamical system
that forced the assumption of asymptotic linearity upon us in the relativistic case.
In particular we can naturally include asymptotically
polytropic behavior as $p\rightarrow\infty$ in the Newtonian case
(cf.~\cite{heiugg03}).

In Newtonian gravity there exist two independent dimensional scales,
space and time, while in relativity, space and time are connected
by the speed of light $c$,
so that only one single dimensional scale remains.
A Newtonian equation of state can be written in the implicit form
\begin{equation}
\label{newtstate}
k_1 p = \phi(\omega)\;, \qquad k_2 \rho = \chi(\omega)\;,
\end{equation}
where $\omega$ is a dimensionless variable and $k_1$ and $k_2$ are constants carrying
the dimension $[p]^{-1}$ and $[\rho]^{-1}$ respectively;
$\phi$ and $\chi$ are
monotone functions in $\omega$ containing any number of dimensionless
parameters. An explicit representation of~(\ref{newtstate}) is
$k_2 \rho = \psi(k_1 p)$.

Since our dynamical system is expressed in terms of purely
dimensionless quantities, the dimensional parameters have to drop
out. Consequently, a single dynamical system must be capable of
describing the entire two-parameter set of equations of
state~(\ref{newtstate}) parameterized by $k_1$ and $k_2$. Indeed,
the relevant function entering the dynamical system in the
Newtonian case is $\Upsilon(\omega)$ given by $\Upsilon(\omega) =
(d\ln\chi/d\ln\omega) (d\ln\phi/d\ln\omega)^{-1}$. Evidently, this
function does not depend on $k_1$ and $k_2$. The same clearly
holds for $f(\omega)=d\ln\phi/d\ln\omega$.

\begin{example}
As an illustrative example, consider the equation of state
$\rho = C_1 p^{\Upsilon_0} + C_2 p^{\Upsilon_1}$. By
rescaling the equation and the two parameters one can write
the equation of state as
$k_2 \rho = (k_1 p)^{\Upsilon_0} + (k_1 p)^{\Upsilon_1}$,
and thus the equation has been brought to the desired form.
The choice $\omega = (k_1 p)^a$ ensures that
$\Upsilon(\omega)$ and therefore also $\Upsilon(\Omega)$ are
independent of $k_1$ and $k_2$, and interpolate
monotonically between $\Upsilon_0$ and $\Upsilon_1$.
In analogy to the detailed discussion in Sec.~\ref{equationsofstate},
the constant $a$ must be chosen sufficiently small so that
a ${\mathcal C}^1$-differentiable dynamical system is obtained.
\end{example}

\begin{remark}({\em Consistent translatory invariance}).
As seen previously, it is the freedom to choose $\omega(p)$
that allows one to treat a one-parameter class of equations of state
simultaneously by one specified dynamical system in the relativistic case.
The freedom we were able to exploit in the Newtonian case
to cover even two-parameter classes of equations of state
reflects itself in the
consistent translatory invariance of the Newtonian dynamical system.
Namely, if ${\bf x}(\xi)$ (${\bf x}(\xi):=(u(\xi),v(\xi),\omega(\xi))$
(recall that $\xi = \ln r$)
is a solution of the Newtonian dynamical system, then so is
${\bf x}_0:={\bf x}(\xi-\xi_0)$,
and moreover, the translated solution gives rise to a perfect fluid
solution associated with an equation of state with the same $\Upsilon(p)$.
In contrast, in the relativistic case this consistency is broken:
if ${\bf x}(\xi)$ is a solution of the
dynamical system~(\ref{uvomeq}), then so is
${\bf x}_0(\xi) = {\bf x}(\xi-\xi_0)$.
However, whereas the solution ${\bf x}(\xi)$ can be consistently
interpreted as a relativistic perfect fluid solution,
associated with an equation of state $\rho(p)$ (determined
by $\Upsilon$ and $\sigma$), this is not the case
with the translated solution. Although the translated
solution satisfies differential equations associated with $\rho(p)$,
the initial data are not consistent with this equation of state.
Accordingly, only one particular solution on every orbit
of~(\ref{uvomeq}), can be interpreted as a relativistic perfect fluid
solution. However, in the compactified
dynamical system~(\ref{UVOmega}) this `defect' is remedied by
the appearance of the freedom in
the new independent variable $\lambda$ (this `cure' could also
have been implemented in the $u,v,\omega$-formulation by defining
$\xi$ through $dr/d\xi=r$, i.e, by letting $\xi= \ln r + \xi_0$, where
$\xi_0$ is an arbitrary constant, instead of setting $\xi = \ln r$,
however, the direct relation between $\xi$ and $r$ is sometimes useful).
The reason behind the difference in the Newtonian and relativistic
cases is the appearance of $\sigma$ in
the relativistic dynamical system; $\sigma$ uniquely determines
the equation of state while $\Upsilon$ only does so up
to a proportionality constant.
\end{remark}

To conclude the remarks on the Newtonian case, we
consider the Newtonian counterpart of the mass-radius formulas~(\ref{AMR}).
These equations simplify
considerably in the Newtonian case.
Starting from the two-parameter family~(\ref{newtstate})
we obtain for the surface potential
$\Phi_N(R)= - G M/R$ the expression
$G M/R = k_1^{-1} k_2 B/s$, and for $R^2$ and $M^2$,
\begin{equation}
R^2 = \frac{k_2^2}{4\pi s^2 k_1\,G}\:A B^{1-n_0}\ ,
\qquad
M^2 = \frac{k_2^4}{4\pi s^4 k_1^3\, G^3}\: A B^{3-n_0}\:.
\end{equation}

\end{appendix}


\end{document}